\def\aa{A\&A}                                     	  % A&A
\def\aas{A\&AS}                                      % A&AS
\def\apj{ApJ}                                      	 % ApJ
\def\apjs{ApJS}                                  	 % ApJS
\def\aj{AJ}                                         	 % AJ
\def\mnras{MNRAS}                                % MNRAS
\def\aap{AAP}                                         % AAP
\def\jrasc{JRASC}                                   % JRASC
\def\physrep{Phys. Rep.}                         % Phys. Rep.
\def\lsim{\lower.5ex\hbox{$\; \buildrel < \over \sim \;$}}
\def\gsim{\lower.5ex\hbox{$\; \buildrel > \over \sim \;$}}
\documentclass{aa}
\usepackage{epsfig}

\begin{document}

\title{Microquasars as sources of positron annihilation radiation}

\author{N.~Guessoum\inst{1}
         \and P.~Jean\inst{2}
         \and N.~Prantzos\inst{3}
       }

    \authorrunning{N.~Guessoum et al.}

   \titlerunning{Microquasars as sources of positron annihilation 
radiation}

\institute{ 
 $^{1}$ American University of Sharjah, College of Arts \& Sciences, 
 Physics Department, PO Box 26666, Sharjah, UAE \\
 $^{2}$ CESR, CNRS/UPS, B.P.~4346, 31028 Toulouse Cedex 4, France \\
  $^{3}$ Institut d'Astrophysique de Paris, 98bis Bd. Arago, 75104 
Paris, France \\
          }

\date{Received  ; accepted }

\abstract{ We consider the production of positrons in microquasars, 
i.e. X-ray binary systems that exhibit jets frequently, but not 
continuously. 
We estimate the production rate of positrons in microquasars, both by 
simple 
energy considerations and in the framework of various proposed 
models. 
We then evaluate the collective emissivity of the annihilation 
radiation 
produced by Galactic microquasars and we find that it might 
constitute a 
substantial contribution to the  annihilation flux measured by 
INTEGRAL/SPI. 
We also discuss the possible spatial distribution of Galactic 
microquasars, 
on the basis of the (scarce) available data and the resulting 
morphology of 
the flux received on Earth. Finally, we consider nearby ``misaligned" 
microquasars, with jets occasionally hitting the atmosphere of the 
companion star; 
these would represent interesting point sources, for which we 
determine the 
annihilation flux and the corresponding light curve, as well as the 
line's 
spectral profile. We discuss the possibility of detection of such 
point 
sources by future instruments.
}

\maketitle


%%%%%%%%%%%%%%%%%%%%%%%%%%%%%%%%%%%%%%%%%%%%%%%%%%%%%%%%
% 1. Introduction
%%%%%%%%%%%%%%%%%%%%%%%%%%%%%%%%%%%%%%%%%%%%%%%%%%%%%%%%

\section{\label{s1}Introduction: Galactic Positron Sources}

Ever since the first detection of the gamma-ray signature (511 keV 
radiation) of the annihilating positrons in the Galaxy in 1970 
(Johnson, Harnden and Haymes, 1972), the problem of the birth, 
propagation, interaction, and 
annihilation of the positrons has been a major topic of astrophysical 
investigation, both observationally and theoretically. Nonetheless, 
some of its fundamental issues, particularly the questions of 
positron origin and propagation, have remained mostly unsolved 
puzzles.

A series of observational campaigns, conducted by 
balloons\footnote{the pioneering Rice University group's campaign, 
the Bell-Sandia group's, GRIS, HEXAGONE} in the first two decades and 
then by satellites\footnote{HEAO 3, CGRO-OSSE, WIND-TGRS, 
INTEGRAL-SPI}, have produced increasingly more precise, 
high-resolution spectra. Recent data have allowed a determination of 
the physical conditions in the interstellar medium where the 
positrons annihilate (Guessoum et al. 2004; Churazov et al. 2005; 
Jean et al. 2006), based on an improved understanding of the various 
physical processes involved (Guessoum, Jean \& Gillard 2005). 
In parallel, considerable  progress has been made in  mapping the 
Galaxy at 511 keV, first by CGRO-OSSE (Purcell et al. 1994; Cheng et 
al. 1997; Purcell et al. 1997; Milne et al. 2000; Milne et al. 2001) 
and, in particular, recently by INTEGRAL-SPI (Kn\"odlseder et al. 
2005). 

Despite that substantial progress, the  origin of the huge amounts of 
positrons produced in the Galaxy ($10^{43}$ e$^+$s$^{-1}$) is still 
eluding us. A large number of potential sources has been proposed 
over the years: cosmic ray interactions with the interstellar medium 
(Ramaty, Stecker \& Misra 1970), pulsars (Sturrock 1971), radioactive 
nuclei produced by explosive nucleosynthesis in supernovae (Clayton 
1973) or novae (Clayton \& Hoyle 1974), compact objects housing 
either neutron stars or black holes (Ramaty \& Lingenfelter 1979), 
matter expelled by red giants (Norgaard 1980) and Wolf-Rayet stars 
(Dearborn \& Blake 1985, Prantzos and Cass\'e 1986), gamma-ray bursts 
(Lingenfelter \& Hueter 1984), (light) dark matter (Rudaz \& Stecker 
1988; Boehm et al. 2004), low-mass X-ray binaries (Prantzos 2004), 
hypernovae (Cass\'e et al. 2004), millisecond pulsars (Wang, Pun, \& 
Cheng 2006), and, most recently, pair production from the collision 
of gamma-ray photons from ``SMall Mass Black Holes" (SMMBHs) and 
X-ray photons from the Galactic Center black hole Sgr A* (Titarchuk 
\& Chardonnet 2006). 

The recent mapping of the Galaxy at 511 keV (Kn\"odlseder et al. 
2005), particularly the very large Bulge/Disc (B/D) ratio inferred, 
has placed rather severe constraints on the positron candidate 
sources. Of the aforementioned potential sources, Type Ia supernovae, 
low-mass X-ray binaries (LMXB's), and dark matter have emerged, for 
various reasons,  as possible (or, rather, not impossible) 
candidates. However, each of these has its own  difficulties: 

1) Although the {\it total positron emissivity of SNIa in the Galaxy} 
matches the observed one, the corresponding {\it bulge} emissivity is 
estimated  (e.g. Prantzos 2004; Diehl, Prantzos \& von Ballmoos 2005) 
to be an order of magnitude lower  than required by SPI observations. 
Even assuming that large systematic uncertainties affect that 
estimate, the expected galactic distribution of SNIa does not 
correspond to the SPI B/D ratio, unless additional assumptions are 
made (e.g. important number of currently undetected SNIa in the 
galactic bulge, or transfer of a large fraction of the disk positrons 
to the bulge through the galactic magnetic field, see Prantzos 2006); 

2) Low-mass X-ray binaries (LMXBs) display a spatial distribution 
considerably peaked toward the Galactic centre region (Grimm, 
Gilfanov \&  Sunyaev 2002), and  their global energetics allow for a 
substantial positron production galaxywide (Prantzos 2004). However, 
the brightest of them (in X-rays) lie in the disc, not in the bulge 
of the Milky Way (Prantzos 2004); if positron emissivity correlates 
with X-ray emissivity (as is commonly assumed), LMXB's should be 
excluded on the basis of inadequate morphology.

3) Dark matter particles as sources of Galactic positrons may also 
suffer from ``morphology'' problems, or at least uncertainties, but 
also from inconsistencies in the framework of particle physics (see 
Ascasibar et al. 2005).

Finally, we must note the important constraint that Beacom \& 
Y\"uksel (2005) have pointed out w.r.t. all models in principle: if 
the positrons are relativistic (E $\gsim 3$ MeV), then the 
annihilation in flight of a fraction of them (which is very large 
when the medium is neutral) will produce continuum photons of high 
energies ($>$ a few MeV) that could be detected by instruments such 
INTEGRAL and COMPTEL. This result places strong constraints on 
models, particularly those that have positrons produced at high 
energies and annihilating ``in flight", i.e. while slowing down.

In the case of LMXBs,  positrons should be produced as e$^+$-e$^-$ 
pairs in the inner regions of their accretion discs. Some of these 
would annihilate locally, but a non-negligible fraction would be 
channeled out by jets---when these exist. In general, positrons would 
reach the ISM relatively far from their source, and they would 
propagate and annihilate, contributing to the diffuse 511 keV 
emission. In special cases, when the jets are strongly inclined 
(``misaligned") toward the plane of the binary system,  positrons 
could periodically hit the atmosphere of the companion star where 
they would produce an annihilation signature, characterized, in 
particular, by its time variability and line profile.

In this paper we consider the possible fate of positrons ejected by 
microquasar jets. We first estimate the rate of positron production 
in jets, by reviewing estimates from other studies and by using 
simple energetics considerations; in that respect, we pay special 
attention to upper limits derived from the INTEGRAL-SPI measurements 
of positron annihilation fluxes from point sources in the Galaxy. 
We then consider two possible consequences: 
1) the collective microquasar contribution to the flux of galactic  
annihilation radiation, which we find to be potentially substantial  
(especially from the central regions); and 2) the annihilation line 
signature of ``misaligned" microquasars. In the latter case, we study 
the fate of positrons in the companion's atmosphere in order to 
compute the photon emissivity, and determine the characteristic 
``light curve" of such a source; we find that interesting 
observational signatures could be obtained from close enough and/or 
strongly active sources.

In Sect.  2 we present a brief overview of X-ray binaries, 
microquasars, and jets. In Sect. 3 we present our estimates of the 
rate of positrons produced by microquasar jets. In Sect. 4 we 
consider the overall contribution of microquasars to the global 
galactic annihilation flux. In Sect.  5 we consider the annihilation 
of positrons in misaligned microquasars. In Sect. 6 we summarize our 
conclusions.

%%%%%%%%%%%%%%%%%%%%%%%%%%%%%%%%%%%%%%%%%%%%%%%%%%%%%%%%
% 2. X-Ray Binaries, Microquasars, and Jets
%%%%%%%%%%%%%%%%%%%%%%%%%%%%%%%%%%%%%%%%%%%%%%%%%%%%%%%%

\section{\label{s2} X-Ray Binaries, Microquasars, and Jets}

X-ray binaries (XRB's) are systems containing a compact object 
(either a neutron star or a stellar-mass black hole) accreting matter 
from a companion star. A total of 280 galactic X-ray binaries are 
currently known: 131 High Mass X-ray Binaries (HMXB's) (Liu, van 
Paradijs \& van den Heuvel 2000), in which the companion star has a 
mass $\geq 5$ solar masses (spectral type O or B) and where the mass 
transfer usually takes place by way of the strong stellar wind; and 
149 Low Mass X-ray Binaries (LMXB's) (Liu, van Paradijs \& van den 
Heuvel 2001), in which the companion has a low mass (spectral type 
later than B), and the mass transfer is carried out by Roche lobe 
overflow. Grimm, Gilfanov \& Sunyaev (2002) have shown that HMXB's 
tend to be distributed along the galactic plane, while LMXB's tend to 
be clustered in low Galactic longitudes. The number of XRB's brighter 
than $2 \times 10^{34}$ erg/s is estimated to  $\sim$700 in the 
Galaxy (Grimm, Gilfanov \& Sunyaev 2002; Paredes 2005).

Among the detected  XRB's, 43 have now been found to exhibit radio 
emission, which is interpreted as synchrotron radiation. It is 
generally believed that the radio emission is evidence of jets. Of 
the 43 objects, 35 are LMXB's and 8 are HMXB's. Moreover, of the 43 
systems, 16 are confirmed cases with resolved jets (Rib\'o 2005, 
Paredes 2005, Chaty 2005). Such systems (a compact object accreting 
from a companion star and ejecting a stream of relativistic 
particles) are known as microquasars. It is likely that all 43  
radio-emitting X-ray binaries (REXB) are microquasars. The population 
of microquasars in the Galaxy is estimated at about 100 (Paredes 
2005), although only 16 are ``officially" known (or confirmed) at 
present. We  note that Pandey et al. (2006a, 2006b) have recently 
reported 4 new REX sources: 
IGR J17091-3624, IGR J17303-0601, IGR J17464-3213, and IGR 
J18406-0539; in addition, two sources, X Nor X-1 and IGR J17418-1212, 
have been presented as microquasars (Klein-Wolt, Homan \& van der 
Klis 2004, and Tsarevsky 2004, respectively). We refer to these six 
sources as ``new tentative microquasars". The catalog of microquasars 
is expected to grow rapidly with the array of instruments and studies 
now investigating these objects (from radio to gamma rays).

Although jets of such systems were first observed in 1979 (in the 
peculiar object SS 433), microquasars were identified and imaged in 
the Galaxy only in 1992 when Mirabel et al. (1992) performed 
high-resolution radio observations of the GC's ``great annihilator" 
1E 1740.7 - 2942 and soon afterwards in GRS 1915+105 (Mirabel \& 
Rodriguez, 1994). 

Recent studies (Falcke, K\"ording \& Markoff 2004; Mirabel 2004; 
Fender, Belloni \& Gallo 2004, 2005) considerably improved our 
understanding of the conditions that lead to the emergence of jets in 
XRBs. First and foremost, a connection has been established between 
X-ray flux and jet formation: jets seem to appear when the accretion 
disc X-ray luminosity is low. Jets are apparently produced when the 
inner disc is replenished; there is a clear general pattern of steady 
jets in the ``low/hard state" of the X-ray (microquasar) sources, 
while no jet is seen in ``the high/soft state". 

A relatively elaborate model of this scenario has been proposed by 
Fender, Belloni \& Gallo (2004, 2005). A relation between the power 
in the jet (when it exists) and the X-ray luminosity of a given 
source is derived: $L_{jet} = A_{steady} L_X^{0.5}$, with 
luminosities  expressed in units of the corresponding Eddington 
luminosity. These authors argue that steady jets are produced when 
the X-ray spectrum of a source hardens beyond a certain value (which 
may be universal or vary somewhat from one source to another). The 
spectrum softens when the X-ray luminosity increases above about 1 \% 
of the Eddington value; when this happens, the jet first increases in 
speed and then quickly gets suppressed and disappears. 

There seems to be an agreement on the above relation between 
L$_{jet}$ and L$_X^{0.5}$, which is supposed to hold for sources 
individually. On the other hand, there seems to be considerable 
uncertainty over the value of $A_{steady}$ (Fender, Maccarone \& van 
Kesteren, 2005); indeed, using various sources (e.g. XTE J1118+480) 
for ``calibration", authors obtain values ranging from $0.006$ as a 
lower limit (Fender, Belloni \& Gallo 2005) to 0.3 (Malzac, Merloni 
\& Fabian, 2004), which Fender, Belloni \& Gallo (2005) take as an 
upper limit. In the case of transient jets, Fender, Maccarone \& van 
Kesteren (2005) conclude that the value of $A_{trans}$ would range 
between 0.04 and 4.0, at least for black hole sources, which tend to 
exhibit jets that are about 10 times more powerful than those of 
neutron star jets, both in the steady and in the transient cases.

Secondly, a general correlation is also found between the velocity of 
the outflow and the X-ray luminosity of the accreting source: 
increases in X-ray luminosity tend to accelerate the jets as long as 
the source remains in the low/hard state. Fender, Belloni \& Gallo 
(2005) further argue that  the velocities of transient jets are 
significantly larger ($\gsim 0.87$ c) than those of steady jets 
($\lsim 0.7$ c). Jets with larger Lorentz factors have also been 
considered, but the standard cases are those presented by Fender, 
Belloni \& Gallo (2005).

The particle content of the jets is among the major unresolved issues 
in the field of microquasars (e.g. Scheck et al. 2002). Observations 
of emission lines in the case of  SS 433  (Marshall, Canizares \& 
Schulz 2002) suggest a substantial baryonic content of its jets. 
Several studies have considered the implications of a hadronic jet 
composition, either for the galactic cosmic ray content (Heinz \& 
Sunyaev 2002, Fender, Maccarone \& van Kesteren 2005) or for the 
synthesis of Li on the surface of the companion star (Butt, Maccarone 
\& Prantzos 2003). Others have considered the gamma-ray and neutrino 
production at the surface of the companion by impinging high-energy 
protons (Romero \& Orellana 2005). On the other hand, arguments for 
e$^+$-e$^-$ pair-dominated jets have been put forward, especially in 
cases involving the extraction of the spin energy of the black hole 
(Celotti \& Blandford 2001). A strong argument in favor of such a 
leptonic composition is the repeated observations of  highly 
polarized jets. Several leptonic microquasar models have been 
proposed in recent times, e.g. Bosch-Ramon and co-workers (2004a, 
2004b, 2005), Dermer \& B\"ottcher (2006); see the  review by Romero 
(2005). Some rather complex models have also been proposed, like 
``two-flow" models (pair beam surrounded  by a mildly relativistic 
e$^-$-p plasma) , or e$^-$-p jets that later get loaded with pairs by 
interactions with high-energy photons (e.g. Scheck et al. 2002 and 
references therein).

In this work we consider microquasar jets channeling positrons (or 
e$^+$-e$^-$ pairs) from the inner regions of the source's accretion 
disc into the ISM or towards the companion star (if the jet is 
``misaligned"). We also note that positrons can also be produced by 
hadrons from the jet colliding with nuclei (e.g. through p - p 
$\rightarrow$ p + n + $\pi^+$), but such processes would contribute 
only negligible amounts of positrons due to the small numbers of 
high-energy hadrons as well as the low values of the relevant cross 
sections.

In order to estimate the flux of the resulting annihilation 
radiation, the rate of positron ejection by the jets must then be 
determined.

%%%%%%%%%%%%%%%%%%%%%%%%%%%%%%%%%%%%%%%%%%%%%%%%%%%%%%%%
% 3. Rate of positrons produced by microquasar jets
%%%%%%%%%%%%%%%%%%%%%%%%%%%%%%%%%%%%%%%%%%%%%%%%%%%%%%%%

\section{\label{s3} Rate of positrons produced by microquasar jets}

The high-temperature, high-density conditions in the inner regions of 
the binary system's accretion disc produce electron-positron pairs 
via $ \gamma + \gamma \longrightarrow e^+ + e^-$ reactions. A 
fraction of them annihilate  close to the compact object, but when 
jets appear, they may channel out a significant number of pairs. A 
few authors have attempted to model and estimate the production and 
ejection of pairs in those conditions.

According to Beloborodov (1999) the pairs are cooled to energies of 1 
-- 10 keV and blown away by soft radiation to form a 
semi-relativistic wind. Depending on the compactness of the source 
($l = L \sigma_T /m_ec^3R$, where $L$ and $R$ are the power and 
radius of the emitting region, and $\sigma_T$ is the Thomson cross 
section), the plasma will form an optically thin or thick atmosphere. 
The density of the outflow and the rate of ejection of electrons and 
positrons depend on the rate of production of pairs (which, in its 
turn, depend on the photon ``seed" spectrum and on the accretion disc 
model), the annihilation rate and the ``escape efficiency" of the 
pairs. Under the assumption of an optically thin pair wind, where the 
pairs escape before they annihilate ($\tau_{esc} < \tau_{ann}$, with 
$\tau_{ann} \sim 1/n_e\sigma_Tc$, where $n_{e}$ is the electron 
density), Beloborodov (1999) shows that the maximum pair luminosity 
is given by
\begin{eqnarray}
 L^{\rm max}_{e^+e^-} = {2 \pi m_e c^3 R \over \sigma_T} \; , 
\end{eqnarray}
\noindent which translates into a rate of pair injection in the jet 
of $\sim 4 \times 10^{41}$s$^{-1}$. A substantial fraction of these 
pairs, perhaps up to 90 \% (as argued by Misra \& Melia 1993), would 
annihilate near the base of the jet, producing  a broad and 
redshifted line (that is rather difficult to detect); as many as 
$\sim 10^{41}$ s$^{-1}$  positrons are then expelled into the ISM 
generally or, occasionally, in the direction of the companion star.

Misra \& Melia (1993) suggest that the intense radiation field is 
responsible for the Compton acceleration of the pairs produced in the 
inner regions of the accretion disc. They find  that a large rate of 
pairs (up to $6 \times 10^{42}$ s$^{-1}$) stream outwards from the 
disc (at velocities of $\sim 0.7$ c), even after 90 \% have 
annihilated near the base. One must note, however, that this large 
rate is obtained  with an accretion rate of $\sim 5 \times 10^{-8} 
M_{\odot}$ yr$^{-1}$, a rate that can normally be attained only in 
episodic outbursts. Indeed, the model of Misra \& Melia (1993) was 
mainly aiming to reproduce the 1E 1740.7 -2942 ``annihilation flare" 
of 1991.

Yamasaki, Takahara \& Kusunose (1999) consider two-temperature 
accretion discs by taking into account the formation of relativistic 
pair outflows, in both AGN and microquasars. They show that in the 
inner regions of the discs, when the mass accretion rate becomes 
larger than about one tenth of the Eddington rate ($1.4 \times 
10^{17} M/M_{\odot}$  g s$^{-1}$ where $M$ is the mass of the compact 
object) or $\sim 2 \times 10^{-10} M_{\odot}$ yr$^{-1}$, most of the 
viscously dissipated energy is converted into the thermal and kinetic 
energy of the electron-positron pairs. They obtain a maximum power of 
the pair outflow of 0.136 $L_{Edd}$ for an accretion rate of 10$^{-9} 
M_{\odot}$ yr$^{-1}$ (assuming $M = 1 M_{\odot}$ and $R = 10 
R_{\odot}$), which translates into a pair ejection rate of $2 \times 
10^{42}$ s$^{-1}$, assuming the jet is leptonic.

Considering the global energetics of microquasar jets in the Galaxy, 
one may note that 
\begin{itemize}
\item{at L$_X$ = 0.5 L$_{Edd}$ one obtains $\dot{N}_{e^+} =2 \times 
10^{42}$ s$^{-1}$ (Yamasaki, Takahara \& Kusunose 1999), which 
in the following estimate we may use as a ``yardstick"; }

\item{steady jets are produced at L$_X$ = 0.01 - 0.1 L$_{Edd}$ 
(Fender, Belloni \& Gallo 2004, 2005);} 

\noindent and
\item{$\dot{N}_{e^+} \propto $ L$_{jet} \propto$ L$_X^a$, where a=0.5 
or 1 (see discussion in Sec. 2) and where the value of the 
proportionality constant A is not needed if one uses the above 
``yardstick";} 
\end{itemize}
one can infer that the positron emissivity of a jet in the steady 
state lies in the range $\dot{N}_{e^+}\sim  4 \times 10^{40} - 9 
\times 10^{41}$ e$^+$ s$^{-1}$, with 10$^{41}$ e$^+$ s$^{-1}$ as a 
reasonable average value.

Finally, in attempting to fit the Nova Muscae 1991 ``annihilation 
flare", Kaiser \& Hannikainen (2002) inferred a positron annihilation 
rate of $2 \times 10^{43}$ s$^{-1}$ in the atmosphere of the 
companion star (assuming a distance to Earth of $\approx 5.5$ kpc). 
They argued that such a high pair yield can be achieved when the 
plasma is ``photon-starved", that is when the number of high-energy 
photons in the disc strongly exceeds that of the soft seed photons 
(see Zdziarski, Coppi \& Lamb 1990).

Despite the wide range in the results obtained in the aforementioned 
studies, one concludes that a reasonable estimate for ``steady-state" 
production and ejection of pairs in normal conditions is $\sim 
10^{41}$ s$^{-1}$.

Taking a different, empirical approach, interesting upper-limit 
constraints to the microquasar positron emissivity may be established 
by way of the recent annihilation radiation measurements by 
INTEGRAL-SPI. Kn\"odleseder et al. (2005) have published 3-$\sigma$ 
flux upper limits for a dozen galactic  microquasars/LMXB's/HMXB's; 
we have completed the data for other sources of interest to us here, 
particularly the ``misaligned" microquasars that we will consider in 
some detail in Sec. 5. Table 1 lists the sources we have considered, 
with the corresponding SPI upper-limits on their 511 keV emission 
flux (taking the sources to be point-like) and inferred rates of 
positron injection, assuming that positrons do not annihilate far 
from the source. The bottom part of the table lists microquasar 
sources, the last two or three being misaligned ones.

%%%%%%%%%%%%%%%%%%%%%%%%%%%%%%%%%%%%%%%%%
% Table - Limits on positron rates from SPI upper limits of some 
% nearby sources
%%%%%%%%%%%%%%%%%%%%%%%%%%%%%%%%%%%%%%%%%

\begin{table*}
\caption{Limits on positron rates from SPI upper limits for XRB 
sources of interest. The positron rates were calculated assuming a 
positronium fraction of 95\%. The bottom part of the table lists 
microquasar sources, the last two or three being misaligned ones.}
\label{tab:spiflux}
\begin{array}[b]{lcccccc}
\noalign{\smallskip}
\hline
\hline
\noalign{\smallskip} 
 \mbox{Source} \quad & \qquad \mbox{Type} \quad & \quad 
\mbox{Distance} \quad & \qquad \mbox{l} \qquad & \qquad \mbox{b} 
\qquad & \quad  \mbox{3$\sigma$ Flux Limit} \quad & \quad 
\mbox{Positron Rate} \\
	 &     &  \mbox{(kpc)} & \mbox{(deg)} & \mbox{(deg)} & \quad 
\mbox{(10$^{-4}$cm$^2$s$^{-1}$)} \quad & \mbox{(e$^+$s$^{-1}$)}   \\
\noalign{\smallskip}
\hline
\noalign{\smallskip}
\mbox{GX 349+2} & \mbox{LMXB} & \mbox{$\lsim 10$} & \mbox{349.1} & 
\mbox{2.75} & \mbox{0.8} & \mbox{$\gsim 1.7 \times 10^{42}$} \\

\mbox{GX 5-1} & \mbox{LMXB} & \mbox{8} & \mbox{5.08} & \mbox{-1.02} & 
\mbox{0.7} & \mbox{$9.4 \times 10^{41}$} \\

\mbox{Nova Muscae} & \mbox{LMXB} & \mbox{3} & \mbox{295.3} & 
\mbox{-7.07} & \mbox{2} & \mbox{$3.8 \times 10^{41}$} \\

\mbox{A 0620-00} & \mbox{LMXB} & \mbox{2} & \mbox{209.96} & 
\mbox{-6.54} & \mbox{3.8} & \mbox{$3.1 \times 10^{41}$} \\

\mbox{Cen X-4} & \mbox{LMXB}  & \mbox{1.2} & \mbox{332.24} & 
\mbox{23.89} & \mbox{1.7} & \mbox{$5.2 \times 10^{40}$} \\

\noalign{\smallskip}
\hline
\noalign{\smallskip}

\mbox{GRS 1915+105} & \mbox{LMXB} & \mbox{12.5} & \mbox{45.37} & 
\mbox{-0.22} & \mbox{1} & \mbox{$3.3 \times 10^{42}$} \\

\mbox{Cir X-1} & \mbox{LMXB}  & \mbox{10} & \mbox{322.12} & 
\mbox{0.04} & \mbox{1.1} & \mbox{$2.3 \times 10^{42}$} \\

\mbox{Cyg X-3} & \mbox{HMXB} & \mbox{9} & \mbox{79.85} & \mbox{0.7} & 
\mbox{1} & \mbox{$1.7 \times 10^{42}$}\\

\mbox{1E 1740.7-2942} & \mbox{LMXB} & \mbox{8.5} & \mbox{359.1} & 
\mbox{-0.11} & \mbox{0.9} & \mbox{$1.4 \times 10^{42}$}\\

\mbox{GRS 1758-258} & \mbox{LMXB}  & \mbox{8.5} & \mbox{4.51} & 
\mbox{-1.36} & \mbox{0.7} & \mbox{$1.1 \times 10^{42}$}\\

\mbox{GX 339} & \mbox{LMXB} & \mbox{$\gsim 8$} & \mbox{0.68} & 
\mbox{-0.22} & \mbox{0.8} & \mbox{$\lsim 10^{42}$}\\

\mbox{SS 433} & \mbox{HMXB} & \mbox{4.8} & \mbox{39.69} & 
\mbox{-2.24} & \mbox{0.9} & \mbox{$4.5 \times 10^{41}$} \\

\mbox{LS 5039} & \mbox{HMXB} & \mbox{2.9} & \mbox{16.88} & 
\mbox{-1.29} & \mbox{0.9} & \mbox{$1.5 \times 10^{41}$} \\

\mbox{Sco X-1}  & \mbox{LMXB}  & \mbox{2.8} & \mbox{359.1} & 
\mbox{23.78} & \mbox{1.5} & \mbox{$2.4 \times 10^{41}$} \\

\mbox{Cyg X-1} & \mbox{HMXB} & \mbox{2.5} & \mbox{71.33} & 
\mbox{3.07} & \mbox{1} & \mbox{$1.4 \times 10^{41}$} \\

\mbox{XTE J1118+480} & \mbox{LMXB} & \mbox{2.5} & \mbox{157.66} & 
\mbox{62.32} & \mbox{4.5} & \mbox{$5.9 \times 10^{41}$} \\

\mbox{LS I +61$^o$ 303} & \mbox{HMXB} & \mbox{2} & \mbox{135.68} & 
\mbox{1.09} & \mbox{3.3} & \mbox{$2.8 \times 10^{41}$}\\

\mbox{IGR J17091-3624} & \mbox{?} & \mbox{8.5?} & \mbox{-10.48} & 
\mbox{2.21} & \mbox{0.7} & \mbox{$1.1 \times 10^{42}$} \\

\mbox{IGR J17303-0601} & \mbox{LMXB} & \mbox{8.5?} & \mbox{17.93} & 
\mbox{-1.61} & \mbox{0.9} & \mbox{$1.4 \times 10^{42}$} \\

\mbox{IGR J17464-3213} & \mbox{LMXB} & \mbox{8.5?} & \mbox{-2.87} & 
\mbox{15.01} & \mbox{0.9} & \mbox{$1.4 \times 10^{42}$} \\

\mbox{IGR J18406-0539} & \mbox{?} & \mbox{8.5?} & \mbox{26.67} & 
\mbox{-0.17} & \mbox{1} & \mbox{$1.5 \times 10^{42}$} \\

\mbox{X Nor X-1} & \mbox{LMXB} & \mbox{8.5?} & \mbox{-23.09} & 
\mbox{0.25} & \mbox{1} & \mbox{$1.6 \times 10^{42}$} \\

\mbox{IGR J17418-1212} & \mbox{?} & \mbox{8.5?} & \mbox{13.93} & 
\mbox{9.41} & \mbox{0.9} & \mbox{$1.4 \times 10^{42}$} \\

\mbox{XTE J1550-564} & \mbox{Misaligned?} & \mbox{5.3} & 
\mbox{-34.12} & \mbox{-1.83} & \mbox{1.1} & \mbox{$6.6 \times 
10^{41}$}\\

\mbox{V4641 Sgr} & \mbox{Misaligned} & \mbox{9.6} & \mbox{6.77} & 
\mbox{-4.79} & \mbox{0.7} & \mbox{$1.4 \times 10^{42}$}\\

\mbox{GRO J1655-40} & \mbox{Misaligned} & \mbox{3.2} & \mbox{-15.02} 
& \mbox{2.46} & \mbox{0.8} & \mbox{$1.7 \times 10^{41}$}\\

\noalign{\smallskip}
  \hline
\end{array}
\end{table*}

%%%%%%%%%%%%%%%%%%%%%%%%%%%%%%%%%%%%%%%%%%%%%%%%%%%%%%%%

It can be seen that the data of the Table imply that upper limits for 
the steady positron annihilation rate  of individual sources are 
always $>$10$^{41}$ e$^+$s$^{-1}$ (albeit with large uncertainties, 
due in particular to uncertain distance estimates), so the 
upper-limit rates inferred from INTEGRAL data do not clash with the 
average value derived in the works cited above.

From the different considerations and estimates presented above, it 
appears that a "canonical average" rate of $\sim 10^{41}$ 
e$^+$s$^{-1}$ ejected by the microquasar jets is not unreasonable. We 
will adopt this value for the determination of expected fluxes from 
individual sources (section 5), but realizing the corresponding large 
uncertainty, we will also consider lower ($10^{40}$ e$^+$s$^{-1}$) 
values for ``weak-episode jets" as well as larger ($10^{42}$ 
e$^+$s$^{-1}$) ones, in ``strong flare" cases. We immediately note 
that should this average, ``canonical value" prove to be reasonably 
correct, the $\sim$100 microquasars that are believed to exist in the 
Galaxy (Paredes 2005) would produce a global annihilation emissivity 
near that measured by SPI (and previous instruments).

Another upper limit on  the collective emission by galactic 
microquasars may be derived by considering the global energetics. 
Indeed, (i) the total luminosity of LMXRBs in the Milky Way is 2 - 3 
10$^{39}$ erg s$^{-1}$ (Grimm, Gilfanov \&  Sunyaev 2002), while (ii) 
the luminosity of $\sim$ 10$^{43}$ e$^+$ s$^{-1}$ observed by 
SPI/INTEGRAL is $\sim$ 10$^{37}$ erg s$^{-1}$, assuming the 
positronic jets are mildly relativistic (i.e. positrons have energies 
$<$1 MeV). Of course, as Grimm, Gilfanov \&  Sunyaev (2002) note, the 
Galactic LMXRB population is dominated by a dozen bright sources (of  
$\sim$10$^{38}$ erg s$^{-1}$ each) lying in the disk. But the 
remaining fraction of $\sim$ 90 \% is highly clustered towards the 
bulge (Fig. 1 in Grimm, Gilfanov \&  Sunyaev 2002), as required by 
the SPI data and noted in Prantzos (2004), while their collective 
luminosity is 2 - 3 10$^{38}$ erg s$^{-1}$, i.e. 20  times larger 
than required to explain the Galactic positron energetics. We note 
that those are precisely the low-luminosity (sub-Eddington) sources 
that may produce jets (see  the discussion in Sec. 2). According to 
estimates by Paredes (2005), based on current microquasar statistics, 
there are about 100  microquasars in the Milky Way, which corresponds 
to about 1/3 of the $\sim$ 300 LMXRBs in our Galaxy as estimated by 
Grimm, Gilfanov \&  Sunyaev (2002). Applying this correction factor 
of 1/3 to the microquasar energetics,  still leaves about 10$^{38}$ 
erg s$^{-1}$ available for positron production in their jets, i.e. 
about 6 times more than required by SPI data. We  note at this point 
that some of the currently observed microquasars result from HMXRBs, 
but their fraction is rather small (less than 20 \%) and does not 
affect the arguments presented here (see next section for a more 
detailed treatment). Moreover, we wish to stress the large 
(order-of-magnitude) uncertainty that exists in our current knowledge 
of the ratio between the energy that goes into positrons or into the 
jet (as a whole) compared to the energy t
hat is radiated, an issue that is further complicated by the unknown 
lepton-to-hadron content ratio of the jet.

The simple estimate made in the previous paragraph implies that the 
ratio between the positron power and the X-ray luminosity of 
microquasars is, on average, about 16 \%. One must recall, however, 
that there is a relation between the two quantities, namely 
$\dot{N}_{e^+} \sim$ L$_{jet} \sim$ L$_X^{1/2}$ (references given 
above), which  leads to more realistic expectations.

Indeed, calculating the derivative of the cumulative X-ray 
luminosity function given by Grimm, Gilfanov \&  Sunyaev (2002) for 
LMXRB's (see Eq. 15 of Grimm, Gilfanov \& Sunyaev 2002) and assuming 
that 1/3 of LMXRBs are microquasar, we derived a differential 
luminosity distribution function for 
microquasars:
\begin{eqnarray}
 dN/dL_X = 2.6 L_X^{-1.26} \; , 
\end{eqnarray}
\noindent where L$_X$ is in Eddington units (1.3$\times$10$^{38}$ 
ergs/s) and relates to L$_{e^+}$ as L$_{e^+} = $ B L$_X^a$, with $a$ 
= 1 for a linear relationship and $a$ = 1/2 for the alternate, more 
realistic relation.

Now if the total positron power in the Galaxy is required to be 
10$^{37}$ erg/s (i.e. 0.1 in Eddington units), and using the above 
equation (Eq. 2) to integrate for the total power, one finds that for 
$a$ = 0.5, B is between 0.9 and 1.6 \% depending on the assumption 
made on the maximum X-ray luminosity of the microquasar (1 or 0.1 
L$_{Edd}$, respectively), which are very reasonable figures.

Furthermore, having estimated B and thus obtained an analytic 
distribution function for the microquasars' positron emissivity, one 
can obtain the total rate of positrons emitted by all microquasars 
(of different luminosities) in the Galaxy by integrating dN/dL$_X$ 
against the luminosities and the rate of positrons emitted by each 
source, as obtained by Yamasaki, Takahara \& Kusunose (1999). Again, 
for the linear relation ($a$ = 1) one obtains a total rate N$_{e^+, 
tot}$ between 0.3 and 1.4 x 10$^{43}$ e$^+$/s (depending on the 
maximum X-ray luminosity of a given microquasar, i.e. 0.1 or 1 
L$_{Edd}$, respectively); however, for the preferred non-linear 
relationship ($a$ = 1/2), one obtains N$_{e^+, tot}$ between 1.8 and 
3.1 x 10$^{43}$ e$^+$/s (microquasar X-ray luminosities of 0.1 and 1 
L$_{Edd}$, respectively), values that are very encouraging in 
considering microquasars as possible contributors to the overall 
positron annihilation flux from the central galactic regions.

%%%%%%%%%%%%%%%%%%%%%%%%%%%%%%%%%%%%%%%%%%%%%%%
% 4. Contribution of microquasars to the galactic positron 
% annhilation radiation
%%%%%%%%%%%%%%%%%%%%%%%%%%%%%%%%%%%%%%%%%%%%%%%

\section{\label{s5} Contribution of microquasars to the galactic 
positron annhilation radiation}

Assuming that $\sim 100$ microquasars exist in the Galaxy (Paredes 
2005), and noting from the locations plot of the 22 microquasars 
presently known/confirmed (Figure 1) that about half of them are in 
the central regions, i.e. $\sim \pm 25$ degrees from the GC, we can 
estimate the flux of annihilation radiation that can be expected from 
such a population of sources with jets ejecting $\approx 10^{41}$ 
e$^+$s$^{-1}$ on average and compare it to the global galactic centre 
annihilation flux reported by Kn\"odlseder et al. (2005). 

The microquasars can have either ``misaligned" (highly inclined with 
respect to the orbital plane) or ``normal" (low inclination) jets. 
The normal microquasars will pour out their positrons into the ISM, 
where the usual 2-3 photon production processes take place (see 
Guessoum, Jean, \& Gillard 2005).
In the case of misaligned  microquasars (about 15-20 \% of the total) 
we assume, following Butt, Maccarone \& Prantzos (2003), that their 
jets hit the companion $\sim$10 \% of the time.
Only 1 of the $\sim 2$ annihilation photons will emerge then, the 
other being absorbed in the companion's atmosphere (see Sec. 5). The 
remaining 90\% of the positrons are thrown into the ISM, where they 
produce 2-3 photons each.

The total flux of annihilating positrons coming from the inner Galaxy 
would be the sum of the flux contributed by the misaligned jets and 
the normal jets:
\begin{eqnarray}
 F = F_{\rm mis} + F_{\rm norm}
\end{eqnarray}
where
\begin{eqnarray}
 F_{\rm norm} = { {f_{\mu Q} N_{\mu Q} (1 - f_{{\rm mis}\mu Q}) 
{\dot N}_{e^+} \over {4 \pi D^2} } \times 2 \times f_{\rm line}}  \; .
\end{eqnarray}
and
\begin{eqnarray}
F_{\rm mis} & = & { f_{\mu Q} N_{\mu Q} f_{{\rm mis}\mu Q} {\dot 
N}_{e^+} \over 
{4 \pi D^2} } \nonumber \\
	& \; & \times [f_{\rm jet-*} + 2\times (1 - f_{\rm jet-*})]\times 
f_{\rm line} \; , 
%\nonumber
%
\end{eqnarray}
\noindent where $N_{\mu Q}$ is the total number of microquasars 
believed to exist in the Galaxy ($\sim 100$), $f_{\mu Q}$ is the 
fraction of $\mu$Q's we assume to be in the inner regions of the 
Galaxy ($\approx 50 \%$), $f_{{\rm mis}\mu Q}$ is the fraction of 
$\mu$Q's assumed to have misaligned jets ($\approx 20 \%$), $f_{\rm 
jet-*}$ is the fraction of the time a misaligned jet will hit the 
companion star's atmosphere ($\approx 10 \%$), ${\dot N}_{e^+}$ is 
the rate of ejection of positrons from  a typical jet ($10^{41}$ 
s$^{-1}$), $f_{\rm line}$ is the fraction of photons emitted with the 
line energy (511 keV) as opposed to continuum (0 -- 511 keV) 
energies, and D is the distance to the Galactic centre (8.5 kpc); 
$f_{\rm line}$ is obtained from ${1 \over 4} f_{\rm Ps} + (1-f_{\rm 
Ps})$, where $f_{\rm Ps}$ is the ``Positronium fraction", i.e. the 
fraction of positrons that annihilate via formation of Positronium 
(the bound e$^+$-e$^-$ system), which has repeatedly been found in 
galactic annihilation radiation measurements to be $\approx 0.95$ 
(see references given in Sec. 1);

With those parameter values, we obtain:
\begin{eqnarray}
 F_{\rm norm} \approx 4 \times 10^{-4} \; {\rm ph \ cm^{-2} \ s^{-1}}
\end{eqnarray}
\begin{eqnarray}
F_{\rm mis} \approx 6 \times 10^{-5} \; {\rm ph \ cm^{-2} \ s^{-1}}
\end{eqnarray}

The total flux  F $\approx 5 \times 10^{-4} \ {\rm ph \ cm^{-2} \  
s^{-1}}$ is within a factor of 2 of the SPI-measured annihilation 
flux ($1.1 \times 10^{-3} \ {\rm ph \ cm^{-2} s^{-1}}$), a result 
that is quite encouraging, considering the uncertainties on different 
parts of the problem (mostly due to our currently limited knowledge 
of microquasar jet energetics).

As noted in Sec. 1, the spatial distribution of the 511 keV flux 
detected by SPI-INTEGRAL puts severe constraints on candidate sources 
of positrons. Although far from complete at present, the available 
sample of known/confirmed microquasars appears encouraging in that 
respect. Figure 1 displays the position and type (LMXB/HMXB) of the 
22 currently known microquasars. 

While fully aware that this list represents only about one fifth of 
the microquasars believed to exist in the Galaxy, we can still 
attempt to determine a Bulge-to-Disc (B/D) ratio of the annihilation  
produced by such sources and compare that with the limits obtained 
from SPI data (Kn\"odlseder et al. 2005), which inferred a rate of 
positron annihilation of $(1.5 \pm 0.1) \times 10^{43}$ e$^+$s$^{-1}$ 
in the bulge and $(0.3 \pm 0.2) \times 10^{43}$ e$^+$s$^{-1}$ in the 
disc. 

Referring to Table 1, which lists the types, positions and distances 
of these microquasars, we note that roughly 80 \% of them are LMXB's, 
about 15 \% of which are in the halo, 35 \% are in the disc, and 
about 50 \% are in the bulge, while 20 \% of the sources are HMXB's, 
$\sim 10$\% of which may be in the bulge and the rest in the disc. 
The canonical average rate of positron production has been taken to 
be about $10^{41}$ e$^+$s$^{-1}$ by a typical jet from an LMXB 
microquasar, while for HMXB microquasars jets are ten times less 
powerful and thus about three times less productive in positrons 
(recall that $ L_{e^+e^-} \sim L_{\rm jet} \sim L_X^{0.5}$). One must 
then take into consideration the confinement probability of positrons 
ejected from these sources; we note that: i) according to Jean et al. 
(2006), positrons produced in the bulge do not escape, they end up 
annihilating in the bulge if their energy is below $\sim$10 MeV, 
which is the case for positrons produced by microquasars, if one 
ignores (as in this basic treatment) internal acceleration 
processes; ii) the scale height of LMXBs in the disk is $\sim$400 
pc, while the gas has a scale height of $\sim$100 pc, so about 50 \% 
of positrons produced by LMXB's there are ejected toward the disk and 
annihilate, while the rest (50 \%) of the positrons are released in 
the halo and either contribute to a diffuse annihilation emission 
(unseen by spectrometers) or propagate following the galactic 
magnetic field lines toward the bulge (Prantzos 2006); iii) the 
escape fraction of positrons produced by HMXB's in the disk is 
unconstrained, and for simplicity we take it to be the same as that 
of positrons from disk LMXB's. With these fractions, the net rate of 
positrons annihilating in the bulge is found to be $\approx 4.1 
\times 10^{42}$ e$^+$s$^{-1}$ (about one third the SPI bulge rate), 
while the net rate of positrons annihilating in the disc is found to 
be $\approx 1.7 \times 10^{42}$ e$^+$s$^{-1}$; this would give a B/D 
ratio of 2.4, which is somewhat smaller
 than the lower SPI limit (B/D)$_{SPI} \geq 3$, indicating that 
positrons produced in the disk are escaping in large(r) fractions 
than assumed here.

One can also turn these estimates around and set limits on positron 
production rates from microquasars, knowing that the net rates in the 
bulge and in the disc cannot exceed those set by SPI as mentioned 
above. (Note that since we are taking a canonical average value of 
positron production rate for all LMXB microquasars and assuming the 
HMXB microquasars to produce only one third as many positrons, for 
luminosity reasons as explained above, the B/D ratio is independent 
of the jet positron rate and cannot help set constraints on it, at 
least in our simple model.) Assuming the ratios of LMXB/HMXB 
microquasars and the escape fractions given in the previous 
paragraph, we find that the SPI limits would be violated if positrons 
are produced at steady rates greater than $\sim 3 \times 10^{41}$ 
e$^+$s$^{-1}$.
Furthermore, we note that our model easily satisfies the Beacom \& 
Y\"uksel (2005) constraint since our positrons have kinetic energies 
between 2.5 keV and 660 keV (jet speeds range from 0.1 c to 0.9 c), 
so their annihilation in flight does not produce continuum emission 
of photons at higher energies.

%%%%%%%% Figure : Microquasars Galaxy Map - GJP-fig1.eps %%%%%%%%
%%%%%%%% Figure : Microquasars Galaxy Map - 5240fig1.eps %%%%%%%%
\begin{figure}[tb]
\includegraphics[width=9.5cm,height=8.5cm]{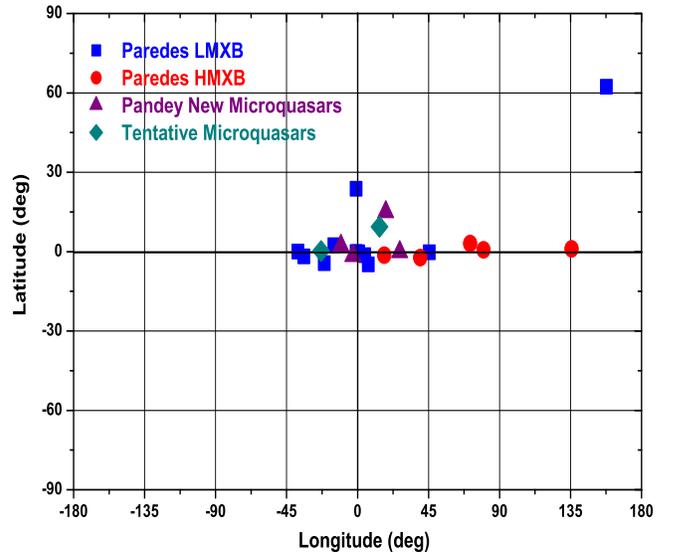}
\caption{Microquasars (position and type) in the Galaxy.}
%\label{fig:mq}
\end{figure}
%%%%%%%%%%%%%%%%%%%%%%%%%%%%%%%%%%%%%%%%%%%

Finally, we note that the line profile that can be expected in this 
scenario is almost exactly the same as the one obtained by SPI and 
analysed recently (Churazov et al. 2005, Jean et al. 2006), since the 
positrons are here poured mostly into the ISM; the small fraction 
ejected from misaligned microquasars that annihilates on the 
companion  star's surface will produce a line of similar profile, as 
the physical conditions (temperature and ionization) in the 
atmospheres of those stars are essentially identical (T $\sim$ 8000 
K, H/He composition, partially ionized) as those in the ISM phases 
where the positrons annihilate on large scales.

%%%%%%%%%%%%%%%%%%%%%%%%%%%%%%%%%%%%%%%%%%%%%%%%
% 5. Annihilation in the Atmosphere of the Companion Star
%%%%%%%%%%%%%%%%%%%%%%%%%%%%%%%%%%%%%%%%%%%%%%%%

\section{\label{s4} Annihilation in the Atmosphere of the Companion 
Star}

In this section we consider in some detail the fate of the jet 
positrons hitting the atmosphere of the companion star, in the case 
of misaligned microquasars. Although this happens in a small number 
of cases ($\sim$15-20\% of all the microquasars, see Sec. 4) and for 
a small fraction of the XRB's period ($\sim$10\%) there are 
potentially interesting gamma ray signatures of such an interaction.

Positrons arrive in the companion star with speeds ranging from 0.1 c 
to 0.9 c (canonical value v$_{jet} \sim$ c/2), i.e. kinetic energies 
of 2.5 keV to 660 keV. They lose energy in the stellar atmosphere 
which has a very sharp density profile (see Fig. 1 in Guessoum \& 
Jean 2002). Using stellar atmosphere models presented in Guessoum \& 
Jean (2002) we estimate the distance  for the energetic positrons to 
slow down below 60 eV to be $\lambda_E \la$ 0.1 g cm$^{-2}$. At this 
energy, positrons are able to form a positronium by charge exchange 
with H atoms. Low energy positrons that do not form a positronium 
quickly thermalize and annihilate in-situ. The maximum atmospheric 
depth at which positrons annihilate ($\lambda_E$) is lower than the 
mean free path of 511 keV photons ($\lambda_{{\rm 511keV}} \sim $6.6 
g cm$^{-2}$) in the atmosphere. Then in these conditions, most ($\ga$ 
0.95$\%$) of the annihilation photons that are emitted upward do 
escape the atmosphere of the companion star. 

This assumption can be considered valid as long as the jet impact 
does not strongly deform the stellar atmosphere. It should be noted, 
however, that the energy deposited by the jets into the star's 
atmosphere ($\sim 10^{36}$ erg s$^{-1}$) is so large as to 
substantially raise the surface temperature (the increase depending 
on various thermodynamical effects) during the impact periods. While 
this phenomenon has no bearing on the annihilation of our positrons, 
it can still lead to some (transitory) observational effects, which 
may be investigated. Indeed, the depth at which positrons 
annihilate depends on the thermodynamic conditions of the gas 
(density, temperature), and using the energy loss rate of positrons 
in a totally ionized medium (see Eq. 3 of Murphy et al. 2005 and 
additional references therein), we estimate that the depth at which 
positrons with initial kinetic energy of 660 keV annihilate is 
$\sim$0.1 g/cm$^2$ for a temperature of 10$^4$ K and $\sim$0.06 
g/cm$^2$ for a temperature of 10$^6$ K. Moreover, at this energy, the 
range of positron can vary also by a factor of $\sim$3.5 with the 
ionization fraction and the abundance (see Fig. 3 of Milne, The \& 
Leising 1999). In any case, the range of 660 keV positron is lower 
than the mean free path of 511 keV photons ($\sim$6.6 g/cm$^2$).

Even if a deformation of the surface does occur, the positrons still 
annihilate at a depth $\sim \lambda_E$ and the upward 511 keV photons 
do escape the atmosphere. However, the impact of the jet produces a 
cavity which may reduce the escaping solid angle of the upward 511 
keV photons. We take this effect into account in an approximate, 
geometric way.

The jet pressure on the stellar atmosphere is $P_{jet} =  L_{{\rm 
jet}}/(v_{{\rm jet}} S) $ , where $L_{{\rm jet}}$ is the power of the 
jet, $v_{{\rm jet}}$ its velocity, and $S$ is the impact surface on 
the secondary. As a first  approximation, we consider that the jet 
stops in the secondary star's atmosphere when the jet pressure is 
equal to the atmospheric gas pressure. At the impact, the jet 
compresses the atmosphere, producing a cavity with a depth that 
depends on the pressure profile of the secondary's atmosphere, the 
jet characteristics (power, velocity and opening angle) and the 
binary separation. For typical values of jet power $L_{{\rm jet}} 
\sim$10$^{36}$ erg s$^{-1}$, velocity $v_{{\rm jet}} \sim$ c/2 and 
opening angle $\theta_{{\rm jet}} \sim$ 1$^{o}$, we obtain $P_{{\rm 
jet}} \sim$ 10$^5$ erg cm$^{-3}$, for a binary star separation of $a$ 
= 2 R$_{\odot}$ and a secondary mass of 0.5 M$_{\odot}$. Using 
the density and the temperature profile models of a 0.5 M$_{\odot}$ 
secondary atmosphere (Guessoum \& Jean 2002), we derived the 
atmospheric gas pressure as a function of the depth and estimated 
cavity depths (below the photosphere) of $\sim$10$^{7}$ cm.

The fraction of upward 511 keV photons that escape the cavity 
can be estimated by calculating the opening angle of the cavity 
(solid angle viewed from the center of the bottom of the cavity to 
the edge of the top of the cavity). Using the same jet-binary 
parameters (the radius of a 0.5 M$_{\odot}$ is $\sim$0.7 R$_{\odot}$) 
and a jet impinging the surface perpendicularly, the radius of the 
jet impact on the secondary atmosphere is 7.9 $\times$ 10$^8$ cm. 
With a depth of 10$^7$ cm, the opening angle is then $\sim$178$^o$, 
yielding a solid angle of 6.2 sr, so the fraction of upward photons 
that escape the cavity is 98$\%$. Considering the most pessimistic 
cases (small binary separation and jet impinging perpendicularly on 
the surface, i.e. high jet pressure), the fraction of upward 511 keV 
photons escaping the cavity is always larger than 70$\%$ (e.g. with a 
separation of a = 1.4$_{\odot}$, the cavity depth is $\sim$10$^8$ cm 
and the escaping fraction is $\sim$77 \%). Therefore, we neglect 
this jet-pressure effect on the secondary atmosphere in the modeling 
of the 511 keV flux presented in the next paragraphs. 

We also checked that the magnetic field of the secondary does not 
affect the jet: we compared the jet pressure with the magnetic 
pressure $P_B = B^2/(4\pi)$; taking a most extreme value for the 
stellar magnetic field $B\sim$10$^3$G we obtain $P_B\sim$ 8 $\times$ 
10$^4$ erg cm$^{-3}$ which is lower than $P_{jet}$.

The upward 511 keV photons that escape the atmosphere may enter the 
incoming jet. They traverse the jet if their mean free path is larger 
than the jet dimension. The minimum mean free path of 511 keV photons 
in the jet material is $\lambda > (n_{e} \sigma_{T})^{-1}$, with 
$n_{e}$ the electron plus positron density in the jet and 
$\sigma_{T}$ the Thompson cross-section. At the secondary surface, 
$n_{e}$ can be approximated by $n_{e} \sim 2 {\dot N}_{e^+}/(v_{{\rm 
jet}} S)$. Assuming a close binary system (e.g. $a$ = 2 
R$_{\odot}$ and R$_{2}$ = 0.7 R$_{\odot}$) to estimate the smallest 
impact area $S$ and therefore the largest density $n_{e}$, we obtain 
$\lambda >$ 6 R$_{\odot}$ (assuming $v_{{\rm jet}} \sim$ c/2 and 
$\theta_{{\rm jet}} \sim$ 1$^{o}$).
This lower limit is already larger than the size of the XRB 
separation. Consequently, the jet itself can be considered as 
quasi-transparent to those gamma photons and the general conclusion 
is that positrons annihilate near the surface, and about half the 
resulting 511 keV photons will be emitted upwards, towards the 
surface and the incoming jet. 

\begin{figure}[tb]
\includegraphics[width=9.0cm,height=6.5cm]{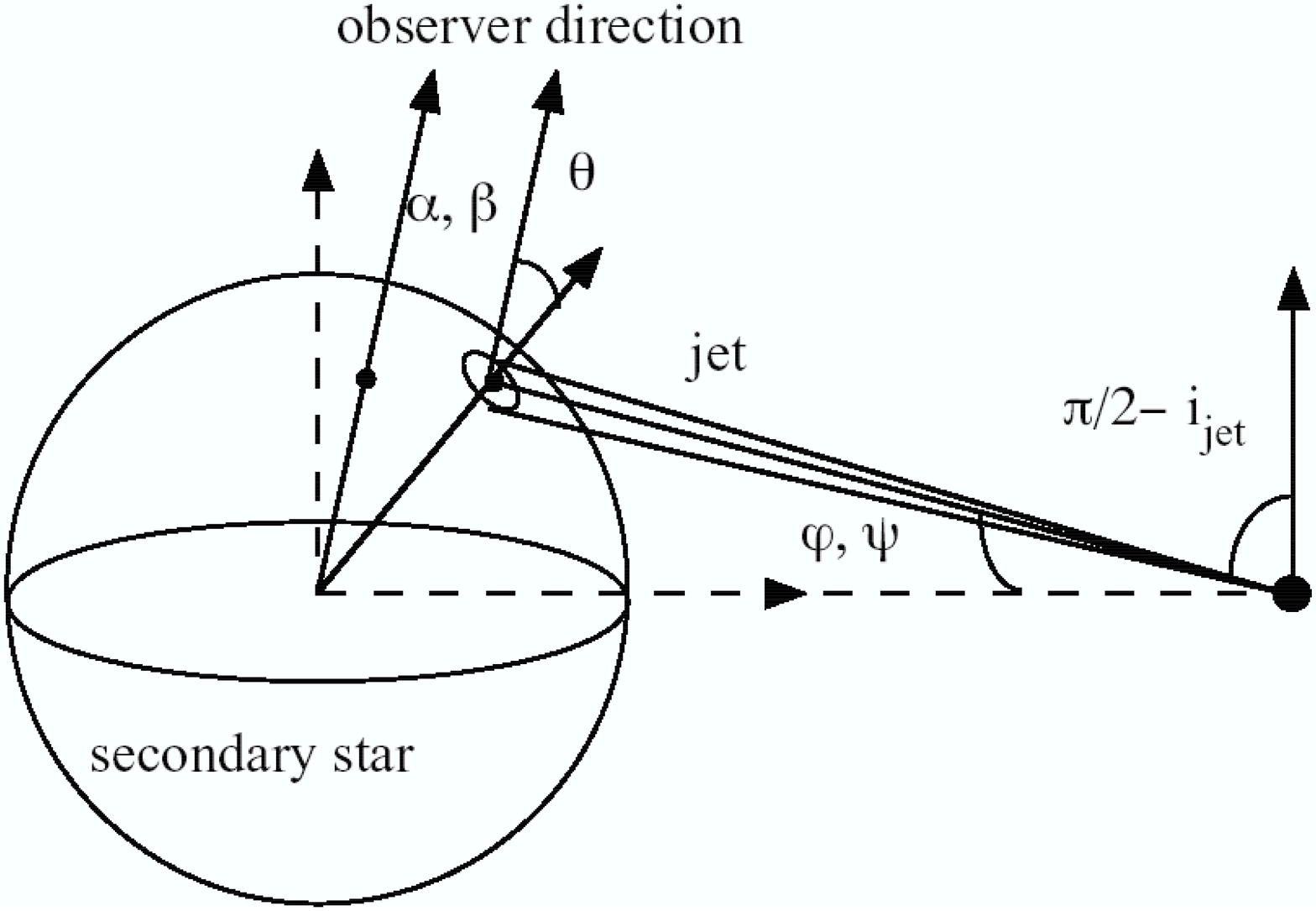}
\caption{Schematic view of the irradiation of the secondary star by a 
jet. The binary system frame is represented by the dashed axes.
\label{fig:geom}}
\end{figure}

The binary system rotates, and the secondary star periodically 
crosses the jet, which slowly precesses. Consequently, the 511 keV 
photon radiation flux received by an observer in a given direction is 
a function of time, and its intensity and time-variability 
characteristics depend on the parameters of the XRB (period, 
inclination, distance, separation and radius of the secondary star) 
and of the jet (inclination with respect to the binary plane, opening 
angle and precession period). The flux at 511 keV can be calculated 
using the following expression:

%%%%%%%%%%%%%%%%%%%%%%%%%%%%%%%%%%%%%%%%%%%%%%%%%
\begin{equation}
     F_{511{\rm keV}}(t) = 2 \times (1-\frac{3}{4}f_{Ps}) \times  
{\dot N}_{e^+}  \times \frac{I(t)}{4\pi d^2}   \label{eq:flux}
\end{equation}
%%%%%%%%%%%%%%%%%%%%%%%%%%%%%%%%%%%%%%%%%%%%%%%%%
\noindent
where $f_{Ps}$ is the positronium fraction, ${\dot N}_{e^+}$ is the 
rate of ejection of positrons from the jet, and $d$ is the distance 
of the binary system to the observer. The function $I(t)$ is the 
fraction of 511 keV photons coming from the atmosphere in the 
direction of the observer. 
This function takes into account the fraction of the jet's solid 
angle (which contains the positrons) that (i) intersects the 
secondary star surface and (ii) is visible by the observer. Its 
expression is given by:

%%%%%%%%%%%%%%%%%%%%%%%%%%%%%%%%%%%%%%%%%%%%%%%%%
\begin{equation}
I(t) = {1 \over {\Delta\Omega_{{\rm jet}}}} \int_{(\varphi, \psi) \in 
D(t)} 
e^{-\frac{\lambda_E(\varphi,\psi)}{\lambda_{511{\rm keV}} 
cos\theta(\varphi,\psi)}} d\Omega(\varphi,\psi)  \label{eq:i(t)}
\end{equation}
%%%%%%%%%%%%%%%%%%%%%%%%%%%%%%%%%%%%%%%%%%%%%%%%%
\noindent
where $\lambda_E$ is the atmospheric depth (in g cm$^{-2}$) at which 
annihilation occurs, and $\lambda_{511{\rm keV}}$ is the mean free 
path of 511 keV photons in the stellar atmosphere. 
$\Delta\Omega_{{\rm jet}}$ is the solid angle of the jet; $\varphi$ 
and $\psi$ give the direction of the jet (see Fig. \ref{fig:geom}); 
and $D(t)$ is the integration domain ($\varphi, \psi \in D(t)$), 
defined as the intersection of the irradiated area and the visible 
area in the direction of the observer (the set of directions 
$\varphi$ and $\psi$ for which $\theta < 90^{o}$ providing the 
observer direction $\alpha$ and $\beta$). 
The irradiation area (or hotspot) is delimited by the intersection of 
the companion surface with the cone of angle $\theta_{{\rm jet}}$/2 
around the jet direction $\varphi_{{\rm jet}}$ and $\psi_{{\rm jet}}$ 
(see Fig. \ref{fig:geom}).
In Equation \ref{eq:i(t)}, $\theta$ is also a function of $\alpha$, 
$\beta$, the XRB separation ($a$) and the secondary radius ($R_{2}$).
Since $I(t)$ is expressed in the XRB frame that is rotating with 
respect to the observer, $\alpha$, $\varphi_{{\rm jet}}$ and 
$\psi_{{\rm jet}}$ vary with time, while $\beta$, the angle between 
the binary system plane and the observer direction, is equal to 
90$^{o} - i$ ($i$ is the ``official'' inclination of the XRB). 

%%%%%%%% Figure : geometry of the jet & the XRB - GJP-fig2.eps  %%%%%
%%%%%%%% Figure : geometry of the jet & the XRB - 5240fig2.eps  %%%%%
%%%%%%%%%%%%%%%%%%%%%%%%%%%%%%%%%%%%%%%%%%%%%%%%%

%%%%% Figure : Instantaneous flux from GRO J1655-40 - GJP-fig3.eps
%%%%% Figure : Instantaneous flux from GRO J1655-40 - 5240fig3.eps
\begin{figure}[tb]
\includegraphics[width=8.5cm,height=6.0cm]{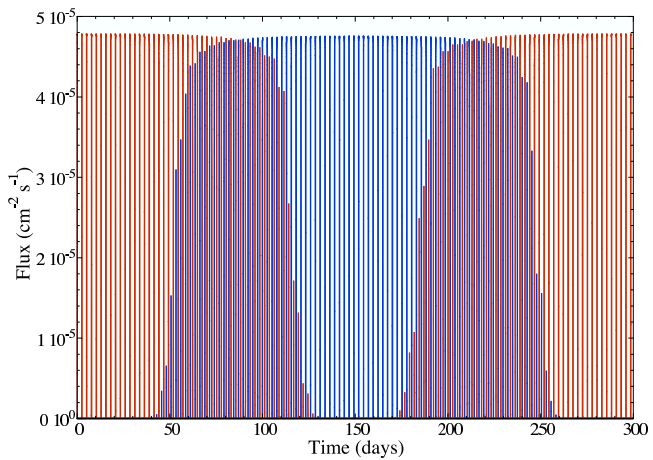}
\caption{Instantaneous 511 keV photon flux induced by the jet (red 
curve) and the counterjet (blue curve) impacts on the secondary of a 
virtual XRB (see text). The jet inclination with respect to the 
binary plane is taken to be $i_{{\rm jet}}$ = 9$^{o}$, its opening 
angle is $\theta_{{\rm jet}}$ = 1$^{o}$; the precession period is 300 
days (see text). 
The positron rate and energy in each jet is ${\dot N}_{e^+}$ = 
10$^{41}$ s$^{-1}$ and $E$ = 660 keV (v$_{{\rm jet}}$ = 0.9 c), 
respectively. 
\label{fig:fluxi}}
\end{figure}
%%%%%%%%%%%%%%%%%%%%%%%%%%%%%%%%%%%%%%%%%%%%%%%%%%%%%%%%

To illustrate the temporal variations of the annihilation flux, we 
calculated the 511 keV light curve for the case of a  
``prototypical misaligned microquasar", one that has similar 
parameters (separation, inclination, primary and secondary 
masses\ldots) as GRO J1655-40, which is one of two XRB's for 
which independent information is available on the jet orientation 
(Narayan \& McClintock 2005). However, we arbitrarily assumed a jet 
inclination of $i_{{\rm jet}}$ = 9$^{o}$ with respect to the binary 
plane. With such an inclination, the jets hit the companion star at 
about half of its upper hemisphere. We also assumed an opening angle 
$\theta_{{\rm jet}}$ = 1$^{o}$ and a precession period of 300 days 
which is the order of magnitude of the precession period of such an 
inclined jet (Kaufman Bernad\'o, Romero \& Mirabel 2002).

Could there be some observational effects on the radio emission 
from the interaction of the jet with the surface of the star? In this 
regard, we first note that pairs in the jets will emit synchrotron 
radiation for most (say 90 \%) of their transit (ejection) time, 
while the rest of the time they spend depositing their energies in 
the companion's atmosphere; this energy is lost first and foremost in 
the form of heat and then, at the end, as gamma rays, which get 
downgraded as X-rays and again as heat. It is thus not expected that 
much radio emission will come out of the jet's collision with the 
companion star's atmosphere, but this issue could be investigated 
further in the future.

\begin{figure}[tb]
\includegraphics[width=8.5cm,height=6.0cm]{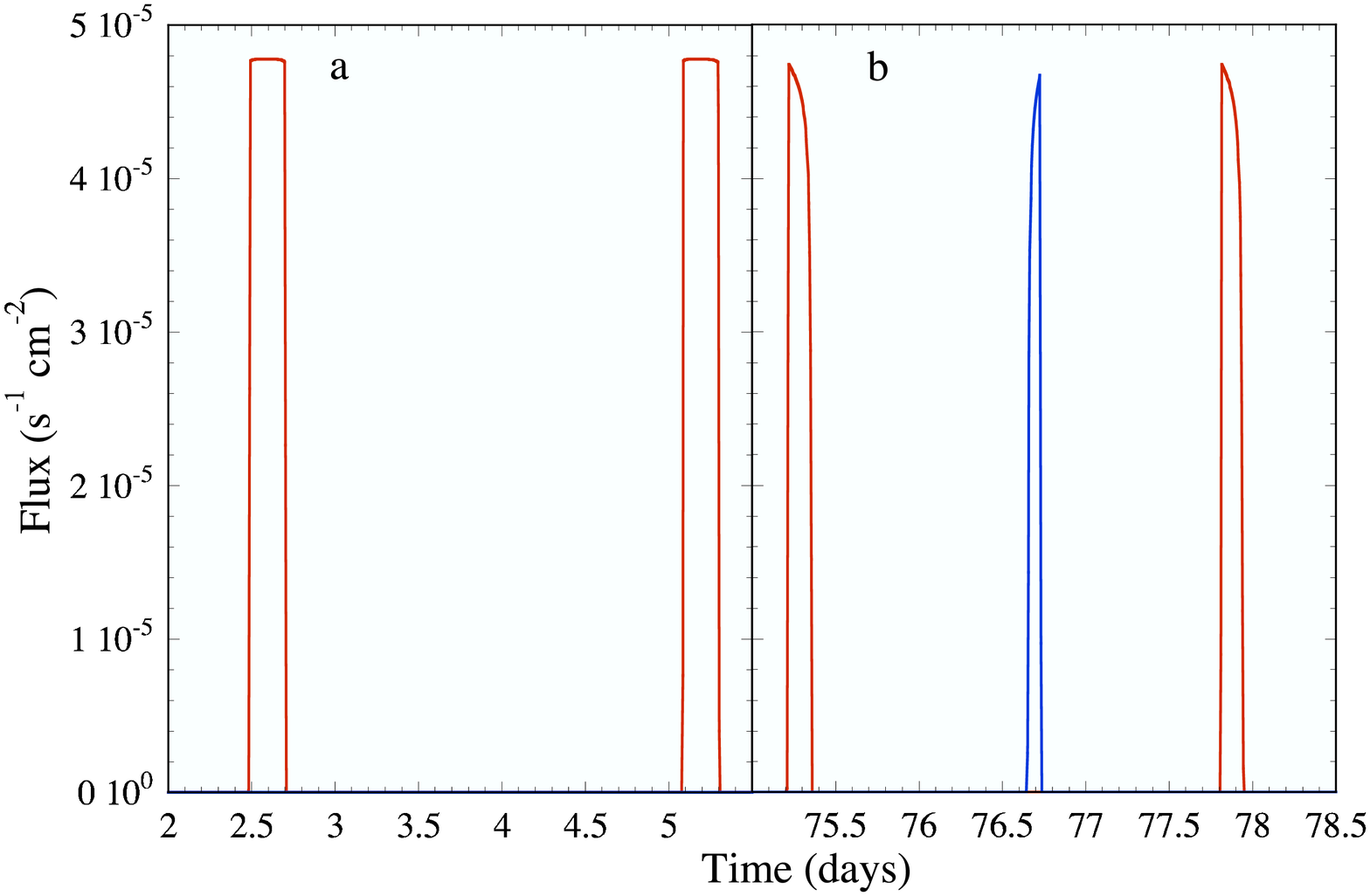}
\caption{Details of the instantaneous 511 keV photon flux induced by 
the jet (red curve) and the counterjet (blue curve) impacts on the 
secondary of a virtual XRB (see caption of Fig. \ref{fig:fluxi}). (a) 
The jet direction is parallel to the observer direction and only the 
hotspot induced by the first jet is visible; (b) the jet direction is 
perpendicular to the observer direction and both hotspots induced by 
the jet and the counterjet are visible. The geometrical 
configurations of the jets and the binary system in these two cases 
are shown in Fig. \ref{fig:rotsys}. 
\label{fig:fluxid}}
\end{figure}

\begin{figure}[tb]
\includegraphics[width=8.5cm,height=9.666cm]{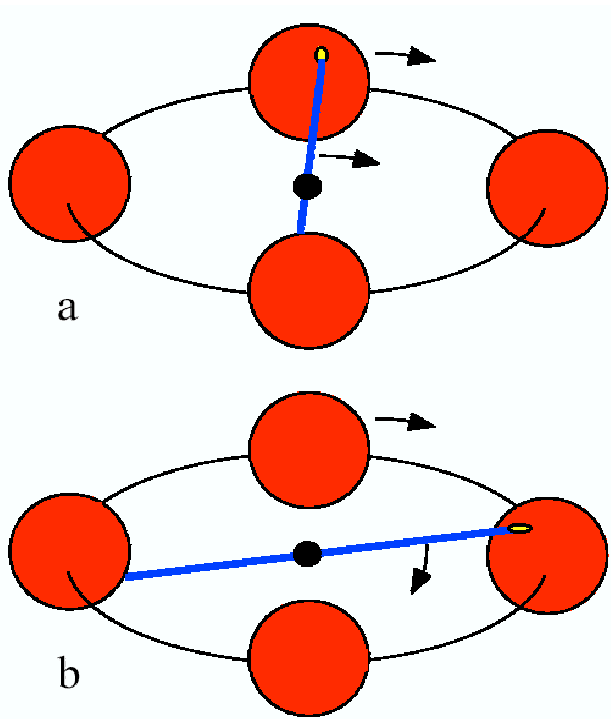}
\caption{Schematic views of the jets (blue) arising from the compact 
object (black), the secondary star (red) presented at four positions 
in its orbit, and the annihilation hotspot (yellow) emitted from the 
secondary star atmosphere, presented at the dates $\sim$ 2 days (a) 
and $\sim$ 75 days (b). The directions of motion of the secondary 
star and jets axis are represented by arrows. The two configurations 
correspond approximately to the cases a and b of Fig. 
\ref{fig:fluxid}. 
\label{fig:rotsys}}
\end{figure}

As for GRO J1655-40, this ``prototypical" microquasar was assumed to 
be at a distance of 3.2 kpc from Earth, to have a period of 2.62 days 
and to be composed of a compact object with a mass of 6.8 M$_{\odot}$ 
separated by 16.6 R$_{\odot}$ from a companion star with a radius of 
5 R$_{\odot}$. The inclination of the binary system is $i$ = 
70$^{o}$. Figure \ref{fig:fluxi} shows the instantaneous 511 keV flux 
emitted by such a microquasar as a function of time. Details of this 
light curve are presented in Figure \ref{fig:fluxid} for two 
different dates: (a) when the direction of the jets lies 
approximately along the line of sight, and in (b) when the direction 
of the jets is approximately perpendicular to the line of sight. The 
geometrical configurations of the jets and the binary system in these 
two cases are shown in Fig. \ref{fig:rotsys}.

%%%%%%%% Figure : Instantaneous flux from GRO J1655-40 - GJP-fig4.eps
%%%%%%%% Figure : Instantaneous flux from GRO J1655-40 - 5240fig4.eps
%%%%%%%%%%%%%%%%%%%%%%%%%%%%%%%%%%%%%%%%%%%%%%%%%%%%%%%%

%%%%%%%% Figure : Schematic views of the jets and XRB - GJP-fig5.eps
%%%%%%%% Figure : Schematic views of the jets and XRB - 5240fig5.eps
%%%%%%%%%%%%%%%%%%%%%%%%%%%%%%%%%%%%%%%%%%%%%%%%%%%%%%%%

\begin{figure}[tb]
\includegraphics[width=8.5cm,height=6.0cm]{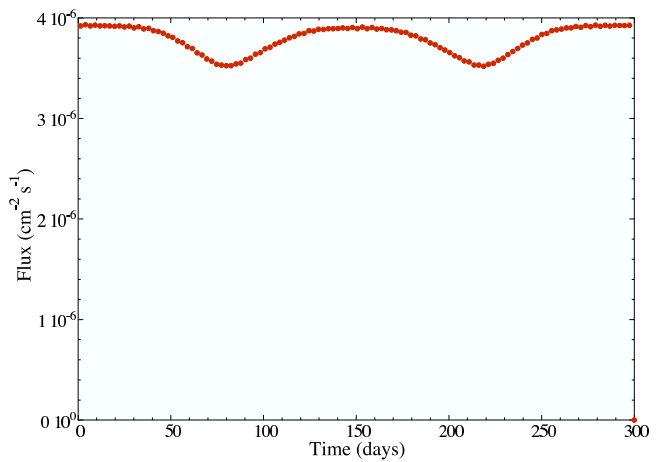}
\caption{Average 511 keV photon flux per XRB period, induced by the 
jet impacts on the secondary of an XRB. The parameters of the 
jets are identical to those of Fig. \ref{fig:fluxi}.
\label{fig:fluxp}}
\end{figure}

Figure \ref{fig:fluxp} shows the time series of the 511 keV flux 
averaged per period of the XRB. The flux is minimum when the jet 
direction is perpendicular to the observer direction, as shown in 
Fig. \ref{fig:rotsys}b. In this configuration, the inclination 
$\theta$ of the direction of the photons emitted by the fraction of 
the hotspot that is visible can be close to 90$^{o}$ during the 
crossing of the secondary. Consequently, a large fraction of these 
photons are absorbed in the atmosphere of the secondary (see Eq. 
\ref{eq:i(t)} and Fig. \ref{fig:fluxid}b). Note that the average flux 
is $\sim$10 times lower than the maximum of the instantaneous flux 
since the companion star crosses the jets during a fraction $\sim$ 
10$\%$ of the period.

%%%%%%%% Figure : Instantaneous flux from GRO J1655-40 - GJP-fig6.eps
%%%%%%%% Figure : Instantaneous flux from GRO J1655-40 - 5240fig6.eps
%%%%%%%%%%%%%%%%%%%%%%%%%%%%%%%%%%%%%%%%%%%%%%%%%%%%%%%%

\begin{figure}[tb]
\includegraphics[width=8.5cm,height=6.0cm]{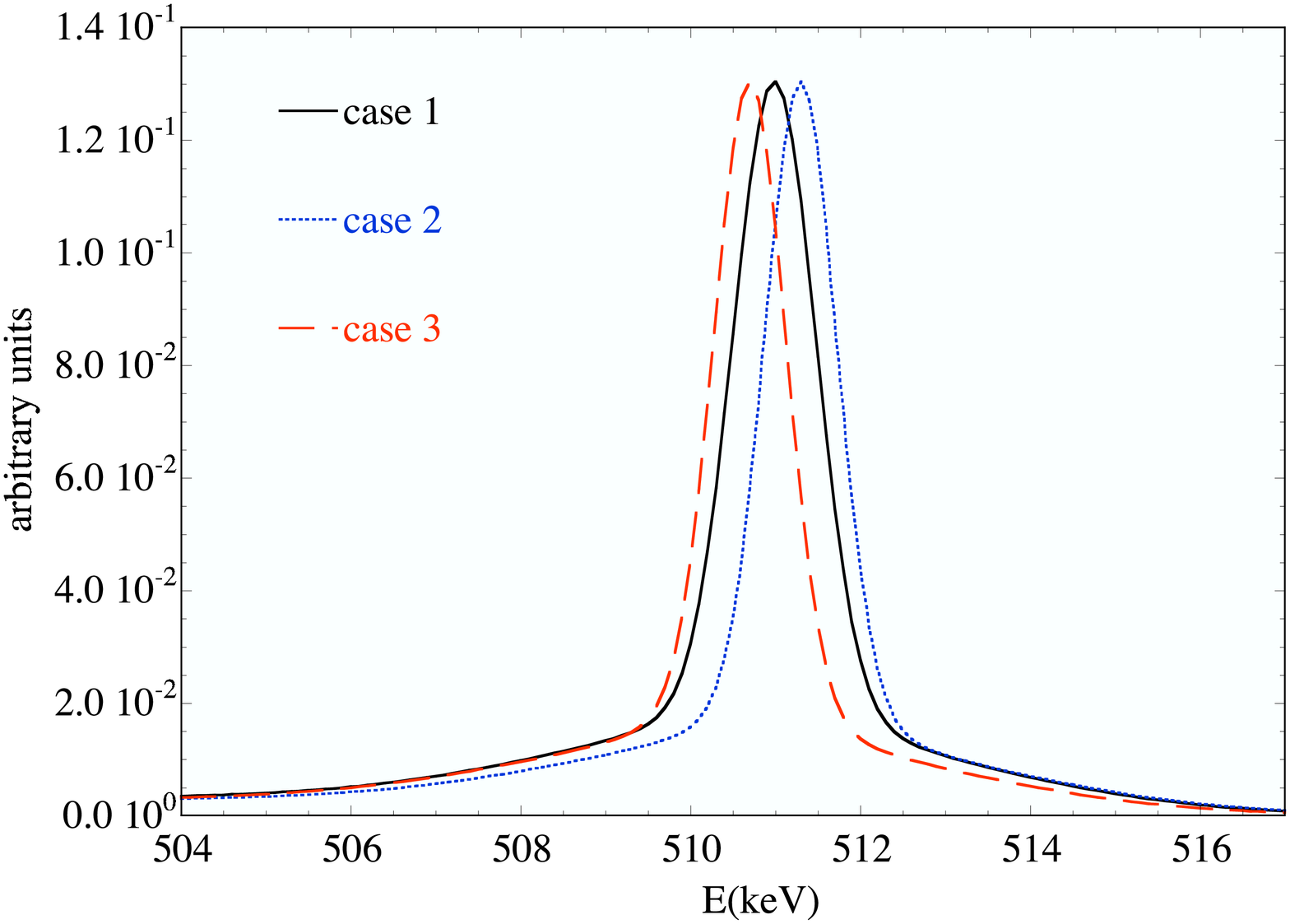}
\caption{Spectra of the annihilation radiation emitted by GRO 
J1655-40. The parameters of the jets are identical to those of Fig. 
\ref{fig:fluxi}. We assumed that the 50$\%$ of positrons annihilates 
with the ionized component of the atmosphere and 50$\%$ with the 
neutral component. case 1: total spectrum integrated over the 
precession period; case 2: spectrum integrated over one period at the 
date 100 days; case 3: spectrum integrated over one period at the 
date 200 days. 
\label{fig:spec}}
\end{figure}

The spectral shape of the annihilation radiation emitted by this 
virtual microquasar was calculated accordingly to the model of 
Guessoum, Jean \& Gillard (2005), which provides annihilation spectra 
as a function of the physical conditions of the gas in which the 
annihilation occurs. We assumed that positrons annihilate in a 
half-ionized and half-neutral plasma with a temperature of 8000 K. 
The rotation of the binary system leads to a shift in the centroid of 
the 511 keV line. This Doppler shift is a function of time and 
depends on the observer direction. We calculated this spectral shift 
using the method described in Jean \& Guessoum (2001). Figure 
\ref{fig:spec} shows the spectra: integrated over the precession 
period (case 1); integrated over one XRB period at t = 100 days (case 
2, see Fig. \ref{fig:fluxi}); integrated over one XRB period at t = 
200 days (case 3). The line is blue-shifted by 0.3 keV in case 2 
because most of the annihilation emission comes from the jet impact 
on the secondary which is moving toward the observer. 
In case 3, most of the annihilation emission comes from the jet 
impact on the secondary which is receding from the observer, thus 
producing a line with a red-shift of 0.3 keV.

%%%%%%%% Figure : Annihilation spectra from GRO J1655-40 - GJP-fig7.eps
%%%%%%%% Figure : Annihilation spectra from GRO J1655-40 - 5240fig7.eps
%%%%%%%%%%%%%%%%%%%%%%%%%%%%%%%%%%%%%%%%%%%%%%%%%%%%%%%%%%%%%%%%%%%%%%%

The results presented in this section show that the measurements of 
the temporal and spectral profiles of the annihilation radiation 
emitted by microquasars with misaligned jets are useful to probe the 
characteristics of the jet. The maximum 511 keV instantaneous flux 
provides the rate of positrons in the jet (e.g. Fig. 
\ref{fig:fluxi}). The duration of the 511 keV emission in one XRB 
period allows one to estimate the opening angle of the jet (e.g. Fig. 
\ref{fig:fluxid}), while the long-term variation of the average flux 
allows for a determination of the precession period of the jet (e.g. 
Fig \ref{fig:fluxp}).  The spectral shape of the 511 keV line, 
do, in principle, allow one to infer the physical conditions of the 
plasma in which positrons annihilate, as Jean et al. (2006) 
demonstrated for the SPI data of positron annihilation in the ISM, 
although one must realize that in more complex situations a host of 
conditions and effects can render the extraction of physical 
information very complicated if not impossible.

Finally, in Table \ref{tab:predic} we present our predicted maximum 
fluxes for the confirmed misaligned microquasars as well as for 
potential candidates, i.e.  microquasars that may be at least 
occasionally misaligned and are not too far away for detection by 
future instruments. We must note that the values given in this 
table are for ``average jets"; ``strong-flare jets" could be 10 times 
more powerful, thus producing 10 times more positrons and thus 
resulting in an annihilation flux 10 times higher (than what is given 
in the table); on the other hand, ``weak episode jets" would give 
positron rates and annihilation fluxes about 10 times lower. Indeed, 
the annihilation flux is directly proportional to the jet power.

%%%%%%%%%%%%%%%%%%%%%%%%%%%%%%%%%%%%%%%%%%%%%%%%%%%%%
% Table - Predicted 511 keV emission fluxes for confirmed or 
% potentially misaligned nearby microquasars
%%%%%%%%%%%%%%%%%%%%%%%%%%%%%%%%%%%%%%%%%%%%%%%%%%%%%

\begin{table*}
\caption{Predicted maximum 511 keV emission fluxes for confirmed or 
potentially misaligned nearby microquasars. Fluxes were 
calculated for ``Average Jet", that is using a positron rate 
of 10$^{41}$ e$^{+}$s$^{-1}$.}
\label{tab:predic}
\begin{array}[b]{lccc}
\noalign{\smallskip}
\hline
\hline
\noalign{\smallskip} 
 \quad\mbox{Source} & \qquad \mbox{Jet} & \qquad \mbox{Distance} & 
\qquad \mbox{Flux for ``Average Jet"} \\
	    		  &     		   &  \qquad \mbox{(kpc)}   & \qquad 
\mbox{(10$^{-4}$cm$^2$s$^{-1}$)} \\
\noalign{\smallskip}
\hline
\noalign{\smallskip}
\mbox{GRO J1655-40} & \quad\mbox{Misaligned} & 3.2 & 0.49  \\

\mbox{V4641 Sgr} & \quad\mbox{Misaligned} & 9.6 & 0.05  \\

\mbox{XTE J1550-564}  & \quad\mbox{Misaligned?}  & 5.3 &  0.18  \\

\mbox{XTE J1118+480} & \quad\mbox{ } & 1.8 & 1.53   \\

\mbox{LS I +61$^o$ 303} & \quad\mbox{ } & 2 &  1.24 \\

\mbox{Cyg X-1}  & \quad\mbox{ }  & 2.5 & 0.79  \\

\mbox{Sco X-1}  & \quad\mbox{ }  & 2.8 &  0.63  \\

\mbox{LS 5039}  & \quad\mbox{ }  & 2.9 & 0.59  \\

\mbox{SS 433} & \quad\mbox{ } & 4.8 &  0.22  \\

\mbox{GRS 1758-258} & \quad\mbox{ }  & 8.5 &  0.07  \\

\mbox{Cyg X-3}  & \quad\mbox{ }  & 9 &  0.06  \\

\mbox{Cir X-1$_2$} & \quad\mbox{ }  & 10 &  0.05  \\

\mbox{GRS 1915+105$_2$} & \quad\mbox{ } & 12.5 &  0.03  \\

\mbox{GX 339-4} & \quad\mbox{ } & \geq 8 &  \leq 0.08  \\

\mbox{XTE J1748-288} & \quad\mbox{ } & \geq 8 &  \leq 0.08  \\

\mbox{1E 1740.7-2942} & \quad\mbox{ } & 8.5 & 0.07  \\

\mbox{IGR J17091-3624} & \quad\mbox{ } & 8.5? &  0.07  \\

\mbox{IGR J17303-0601} & \quad\mbox{ } & 8.5? &  0.07  \\

\mbox{IGR J17464-3213} & \quad\mbox{ } & 8.5? &  0.07  \\

\mbox{IGR J18406-0539} & \quad\mbox{ } & 8.5? &  0.07  \\

\noalign{\smallskip}
\hline
\end{array}
\end{table*}

%%%%%%%%%%%%%%%%%%%%%%%%%%%%%%%%%%%%%%%%%%%%%%%%%%%%%%%%

%%%%%%%%%%%%%%%%%%%%%%%%%%%%%%%%%%%%%%%%%%
% 6.	Conclusions
%%%%%%%%%%%%%%%%%%%%%%%%%%%%%%%%%%%%%%%%%%

\section{\label{s6} Summary}

The origin of the Galactic positron annihilation radiation is a major 
issue in high energy astrophysics at present. In this work we assess 
the potential of Galactic microquasars (XRBs that exhibit jets in an 
intermittent way) as sources of positrons. 

Among the rapidly growing body of data on microquasars, we present in 
Sec. 2 the main features, those of relevance to our study. The 
correlation between the power of the jet and the X-ray luminosity of 
the compact object is the most important of these features. We 
stress, in particular, that the ratio between the two, often assumed 
to be small, is highly uncertain; moreover, the content of the jets, 
leptonic or baryonic (i.e. electron-positron pairs vs. protons and 
pions), is unknown at present. Some of the implications of a baryonic 
content have been explored elsewhere (Butt, Maccarone \& Prantzos 
2003, Romero \& Orelana 2005); we explore here the consequences of a 
leptonic content.

In Sec. 3 we evaluate the rate of positron ejection by the 
microquasar jets, based on various models proposed in the literature, 
but also on simple energetic arguments relating the total power of 
the positrons in the jets to the estimated total X-ray luminosity of 
the ``low luminosity - hard spectrum" galactic LMXRBs.
We find that a value of 10$^{41}$ e$^+$s$^{-1}$ could be considered 
as a ``canonical average'', albeit with large (and difficult to 
evaluate) uncertainties. 

In Sec. 4 we estimate the total positron production rate by the 
collective emission of Galactic microquasars and find it smaller 
(about one third) than what is inferred from SPI/INTEGRAL 
observations of 511 keV radiation. The spatial morphology of the 
corresponding flux received on Earth depends on the assumed large 
scale distribution of Galactic microquasars, which is very poorly 
known at present; we find that the distribution of the currently 
available (incomplete) sample appears encouraging, in that respect. 
(We should also note that a large enough B/D ratio would be obtained 
with any source distributions that congregate in the bulge and/or 
suppress the annihilation of their positrons in the disk.) Finally, 
we constrain the rate of production of positrons by microquasars on 
the basis of the SPI flux results: we find that the SPI limits would 
be violated if positrons are produced at steady rates greater than 
$\sim 3 \times 10^{41}$ e$^+$s$^{-1}$.

Finally, in Sec. 5 we present an investigation of the physics of 
``misaligned'' microquasars, where the jet impinges on the companion 
star. We find that about half of the annihilation photons should, in 
general, escape the stellar atmosphere. We compute fluxes for a list 
of sources (see Table \ref{tab:predic}), a few of them known to be 
misaligned, and some of the others may turn out to be so as well.
The expected annihilation fluxes are found to be low, however, and 
difficult to measure with the spectrometer SPI due to the distance of 
most of the ``misaligned'' microquasars.
Future instruments such as a gamma-ray lens (von Ballmoos et al. 
2004) or a Compton telescope (Boggs \& Jean 2001) would have the 
sensitivity required to measure these fluxes. Due to the rotation of 
the binary system, the secondary star periodically crosses the jet, 
leading to a ``periodic'' and a Doppler-shifted annihilation 
emission. Due to the precession of the jet, we expect to observe a 
long-term modulation of the emission. A detection of a point source 
at 511 keV from a ``misaligned'' microquasar, with such temporal and 
spectral signatures would confirm the presence of positrons in the 
jet. Moreover, such a measurement would allow for a determination of 
the rate of positrons channeled by the jet and the characteristics 
(opening angle, precession period) of the jet.

In summary, we have shown in this study that microquasar jets may 
constitute important sources of annihilation radiation, both diffuse 
(for most of them) or point sources (in the case of misaligned ones).

%%%%%%%%%%%%%%%%%%%%%%%%%%%%%%%%%%%%%%%%%%%%%%%%%%%%%%%%
% Acknowledgements
%%%%%%%%%%%%%%%%%%%%%%%%%%%%%%%%%%%%%%%%%%%%%%%%%%%%%%%%

\begin{acknowledgements}

We wish to acknowledge helpful discussions with Didier Barret and 
Alexandre Marcowith. We are grateful to Mamta Pandey for bringing to 
our notice the four new microquasars that she and her collaborators 
have recently reported. And we thank the anonymous referee for 
several useful suggestions and requests for clarifications, which has 
led to a definite improvement of the paper.

\end{acknowledgements}

\def\aa{A\&A}                                     	  % A&A
\def\aas{A\&AS}                                      % A&AS
\def\apj{ApJ}                                      	 % ApJ
\def\apjs{ApJS}                                  	 % ApJS
\def\aj{AJ}                                         	 % AJ
\def\mnras{MNRAS}                                % MNRAS
\def\aap{AAP}                                         % AAP
\def\jrasc{JRASC}                                   % JRASC
\def\physrep{Phys. Rep.}                         % Phys. Rep.
\def\lsim{\lower.5ex\hbox{$\; \buildrel < \over \sim \;$}}
\def\gsim{\lower.5ex\hbox{$\; \buildrel > \over \sim \;$}}
%
%
%%%%%%%%%%%%%%%%%%%%%%%%%%%%%%%%%%%%%%%%%%%%%%%%%%%
% Bibliography
%%%%%%%%%%%%%%%%%%%%%%%%%%%%%%%%%%%%%%%%%%%%%%%%%%%


\begin{thebibliography}{}

\bibitem[{Ascasibar} (2005)]{Asc05}
Ascasibar, Y. et al., 2005, MNRAS, 368, 1695

\bibitem[{Beloborodov} (1999)]{Bel99}
Beloborodov, A. M. 1999, MNRAS, 305, 181

\bibitem[{Beacom} (2005)]{BY05}
Beacom, J. F., \& Y\"uksel, H., 2005 (astro-ph/0512411)

\bibitem[{Boehm} (2004)]{Boe04}
Boehm, C. et al., 2004, Phys. Rev. Letters, 92, 1301

\bibitem[{Boggs \& Jean} (2001)]{BJ01}
Boggs, S.E. \& Jean P., 2001, A\&A, 366, 1126B

\bibitem[{Bosch-Ramon} (2004)]{Bos04a}
Bosch-Ramon, V., \& Paredes, J. M., 2004, A \& A, 417, 1075

\bibitem[{Bosch-Ramon} (2005)]{Bos05}
Bosch-Ramon, V., Romero, G. E., Paredes, J. M., 2005, A \& A, 429, 267

\bibitem[{Bosch-Ramon} (2004)]{Bos04b}
Bosch-Ramon, V., \& Paredes, J. M., 2004, A \& A, 425, 1069

\bibitem[{BMP} (2003)]{BMP0385}
Butt, Y. M., Maccarone T. J., Prantzos N., 2003, ApJ, 587, 748

\bibitem[{Cass\'e} (2004)]{Cas04}
Cass\'e, M., Cordier B., Paul J., \& Schanne S., 2004, ApJ, 602, 17

\bibitem[{Celotti} (2001)]{Cel01}
Celotti, A. \& Blandford, R. D., 2001, in Proc. of the ESO Workshop 
in Honour of Riccardo Giacconi, ESO Astrophysics Symposia, ed. Kaper 
L., van den Heuvel E.P.J., Woudt P.A., Springer-Verlag, p. 206

\bibitem[{Chaty} (2005)]{Cha05}
Chaty, S. 2005, in Proc. of Rencontres de Moriond, Very High Energy 
Phenomena in the Universe, La Thuile, Italy (March 12-19, 2005), 
(astro-ph/0506008)

\bibitem[{Cheng} (1997)]{Che97}
Cheng, L. X. et al., 1997, ApJ, 481, L43

\bibitem[{Churazov} (2005)]{Chu05}
Churazov, E., Sunyaev R., Sazonov S., Revnivtsev M., \& Varshalovich 
D., 2005, MNRAS, 357, 1377

\bibitem[{Clayton} (1973)]{Cla73}
Clayton, D. D., 1973, Nature Phys. Sci., 244, 137

\bibitem[{Clayton} (1974)]{Cla74}
Clayton, D. D., Hoyle F., 1974, ApJ, 187, 101

\bibitem[{DB} (1985)]{DB85}
Dearborn, D. S. P., Blake J. B., 1985, ApJ, 288, 21

\bibitem[{Dermer-B\"ottcher} (2006)]{DB06}
Dermer, C. D., \& B\"ottcher, M., 2006, ApJ, 643, 1081

\bibitem[{Diehl} (2005)]{Die05}
Diehl, R., Prantzos N., von Ballmoos P., 2005, Nucl.Phys.A, in press 
(astro-ph/0502324)

\bibitem[{FKM} (2004)]{FKM04}
Falcke, H., K\"ording E., Markoff S., 2004, A \& A, 414, 895

\bibitem[{FBG1} (2004)]{FBG04}
Fender, R. P., Belloni T. M., Gallo E., 2004, MNRAS, 355, 1105

\bibitem[{FBG2} (2005)]{FBG05}
Fender, R. P., Belloni T. M., \& Gallo E., 2005, Ap\&SS, 300, 1

\bibitem[{FMK} (2005)]{FMK05}
Fender, R. P., Maccarone T. J., van Kesteren Z., 2005, MNRAS, 360, 
1085

\bibitem[{GGS} (2002)]{GGS02}
Grimm, H. J., Gilfanov M., Sunyaev R., 2002, A \& A, 391, 923

\bibitem[{GJ} (2002)]{GJ02}
Guessoum, N., Jean P., 2002, A \& A, 396, 157

\bibitem[{Guessoum} (2004)]{Gue04}
Guessoum, N., Jean P., Knodlseder J., Lonjou V., von Ballmoos P., 
Weidenspointner G., 2004, Proc. 5th INTEGRAL workshop (Munich), ESA 
SP-552, p. 57

\bibitem[{GJG} (2005)]{GJG05}
Guessoum, N., Jean P., Gillard W., 2005, A \& A, 436, 171

\bibitem[{HS} (2002)]{HS02}
Heinz, S., Sunyaev R., 2002, A \& A, 390., 751

\bibitem[{JG} (2001)]{JG01}
Jean, P., Guessoum N., 2001, A \& A, 378, 509

\bibitem[{Jean} (2006)]{J06}
Jean, P., Knodlseder J., Gillard W., Guessoum N., Ferriere K., 
Marcowith A., Lonjou V., Roques J.-P., 2006, A \& A, 445, 579

\bibitem[{JHH} (1972)]{JHH72}
Johnson, W. N. III, Harnden F. R. Jr., Haymes R. C., 1972, ApJ, 172L, 
1

\bibitem[{KH} (2002)]{KH02}
Kaiser, C. R. \& Hannikainen D. C., 2002, MNRAS, 330, 225

\bibitem[{KBRM} (2002)]{KBRM02}
Kaufman Bernad\'o, M. M., Romero G. E., Mirabel I.F., 2002, A \& A, 
385, L10

\bibitem[{KHK} (2004)]{KHK04}
Klein-Wolt, M., Homan J., van der Klis M., 2004, Nuclear Physics B 
(Proc. Suppl.), 132, 381

\bibitem[{Knodlseder} (2005)]{K055}
Kn\"odlseder, et al., 2005, A \& A, 441, 518 

\bibitem[{Lingen} (1984)]{LH04}
Lingenfelter, R. E., Hueter G. J., 1984, in High-Energy Transients in 
Astrophysics, ed. S. E. Woosley, AIP Conference Proceedings, p. 558

\bibitem[{Liu1} (2000)]{Liu00}
Liu, Q. Z., van Paradijs J., van den Heuvel E. P. J., 2000, A \& A S, 
147, 25

\bibitem[{Liu2} (2001)]{Liu01}
Liu, Q. Z., van Paradijs J., van den Heuvel E. P. J., 2001, A \& A, 
368, 1021

\bibitem[{Marshall} (2002)]{Mar02}
Marshall, H. L., Canizares C. R., Schulz N. S. 2002, ApJ, 564, 941

\bibitem[{Meier} (1996)]{Mei96}
Meier, D., 1996, ApJ 459, 185

\bibitem[{Milne1} (2000)]{Mil00}
Milne, P. A., Kurfess J. D., Kinzer R. L. et al. 2000, AIP Conference 
Proceedings, 510, 21

\bibitem[{Milne01} (2001)]{Mil01}
Milne, P. A., Kurfess, J. D., Kinzer, R. L., Leising M. D., Dixon D. 
D., 2001, AIP Conference Proceedings, 587, 11

\bibitem[{Mirabel1} (1992)]{Mir92}
Mirabel, I. F., Rodriguez L. F., Cordier B., Paul J., Lebrun, F., 
1992, Nature, 358, 215

\bibitem[{Mirabel2} (1994)]{Mir94}
Mirabel, I. F., Rodriguez L. F., 1994, Nature, 371, 46

\bibitem[{Mirabel3} (1996)]{Mir96}
Mirabel I. F., Rodriguez L. F., Chaty S., 1996, ApJ, 472L, 111

\bibitem[{Mirabel4} (2004)]{Mir04}
Mirabel, I. F., 2004, Proc. 5th INTEGRAL workshop (Munich), ESA 
SP-552, p. 175

\bibitem[{MM} (1993)]{MM93}
Misra, R., Melia M., 1993, ApJ, 419L, 25

\bibitem[{Murphy} (2005)]{Mur05}
Murphy R.J., Share, G.H., Skibo, J.G. \& Kozlovsky, B. 2005, ApJSS, 
161, 495

\bibitem[{MTL} (1999)]{MTL99}
Milne, P.A., The, L.-S. and Leising, M.D. 1999, ApJSS, 124, 503

\bibitem[{NM} (2005)]{NM05}
Narayan, R., McClintock J.E., 2005, ApJ, 623, 1017

\bibitem[{Norgaard} (1980)]{Nor80}
Norgaard, H., 1980, ApJ, 236, 895

\bibitem[{Pandey} (2006a)]{Pan06a}
Pandey, M., et al. 2006, A \& A, 446, 471

\bibitem[{Pandey} (2006)]{Pan06b}
Pandey, M., et al. 2006, A \& A, 447, 525

\bibitem[{Paredes} (2005)]{Par05}
Paredes, J. M., 2005, Chin. J. Astron. Astrophys., Vol. 5 Suppl., 121

\bibitem[{Prantzos} (2004)]{Pra04}
Prantzos, N., 2004, Proc. 5th INTEGRAL workshop (Munich), ESA SP-552, 
p. 15

\bibitem[{Prantzos} (2006)]{Pra04}
Prantzos, N., 2006, A \& A, 449, 869

\bibitem[{Prantzos86}]{Pra86}
Prantzos, N., Cass\'e, M., 1986, ApJ 307, 324

\bibitem[{Purcell1} (1994)]{Pur94}
Purcell, W. R. et al., 1994, in Proc. Second Compton Symp., eds. C. 
Fichtel, N. Gehrels, and J. Norris (New York: AIP), 403

\bibitem[{Purcell2} (1997)]{Pur97}
Purcell, W. R. et al. 1997, ApJ, 491, 725

\bibitem[{Ramaty1} (1970)]{Ram70}
Ramaty, R., Stecker F. W., Misra, D., 1970, J. Geophys. Res., 75, 1141

\bibitem[{Ramaty2} (1979)]{Ram79}
Ramaty, R., Lingenfelter R. E., 1979, Nature, 278, 127

\bibitem[{Ribo} (2005)]{Rib05}
Rib\`o, M., 2005, in ASP Conference Series: ``Future Directions in 
High Resolution Astronomy: A Celebration of the 10th Anniversary of 
the VLBA", J. D. Romney \& M. J. Reid (eds.), Vol. 340, p. 269

\bibitem[{RO} (2005)]{Rom05}
Romero, G. E. \& Orellana, M. 2005, A \& A, 436, 237

\bibitem[{Rudaz} (1988)]{Rud88}
Rudaz, S., Stecker, F. W., 1988, ApJ, 325, 16

\bibitem[{Scheck} (2002)]{Sch02}
Scheck, L., Aloy M. A., Mar\'i, J. M., G\'omez J. L., MŸller E., 
2002, MNRAS, 331, 615

\bibitem[{Sturrock} (1971)]{Stu71}
Sturrock, P. A., 1971, ApJ, 164, 529

\bibitem[{Titarchuk } (2006)]{TC06}
Titarchuk, L., \& Chardonnet, P., 2006, ApJ, 641, 293

\bibitem[{Tsarevsky} (2004)]{Tsa04}
Tsarevsky, G. S. 2004, The Astronomer's Telegram, \# 239

\bibitem[{von Ballmoos et al.} (2004)]{PvB04}
von Ballmoos, P. et al., 2004, SPIE, 5168, 482

\bibitem[{Wang} (2006)]{Wan06}
Wang, W., Pun C. S. J., Cheng K. S., 2006, A \& A, 446, 943

\bibitem[{Yamasaki} (1999)]{Yam99}
Yamasaki, T., Takahara F., Kusunose, M., 1999, ApJ, 523L, 21

\bibitem[{Zdziarski} (1990)]{Zdz90}
Zdziarski, A. A., Coppi P. S., Lamb D. Q., 1990, ApJ, 357, 149


\end{thebibliography}
\end{document}